\documentclass[aip,jcp,twocolumn,reprint,groupedaddress]{revtex4-1}

\usepackage{amsmath}
\usepackage{amssymb}
\usepackage{bm}
\usepackage{changes}
\usepackage{graphicx}

\newcommand{\quotes}[1]{``#1''}

\setcitestyle{super}  

\begin{document}

\title{The Density Matrix Renormalization Group in Chemistry and Molecular Physics: Recent Developments and New Challenges.}
\date{7 December, 2019}
\author{Alberto Baiardi}
\author{Markus Reiher}
\thanks{Corresponding author, E-mail: markus.reiher@phys.chem.ethz.ch}
\affiliation{ETH Z\"urich, Laboratorium f\"ur Physikalische Chemie, Vladimir-Prelog-Weg 2, 8093 Z\"urich, Switzerland}

\begin{abstract}
In the past two decades, the density matrix renormalization group (DMRG) has emerged as an innovative new method in quantum chemistry relying on a theoretical framework very different from that of traditional electronic structure approaches. The development of the quantum chemical DMRG has been remarkably fast: it has already become one of the reference approaches for large-scale multiconfigurational calculations.
This perspective discusses the major features of DMRG, highlighting its strengths and weaknesses also in comparison to other novel approaches. The method is presented following its historical development, starting from its original formulation up to its most recent applications. Possible routes to recover dynamical correlation are discussed in detail. Emerging new fields of applications of DMRG are explored, such as its time-dependent formulation and the application to vibrational spectroscopy.
\end{abstract}

\maketitle

\section{Introduction}
\label{sec:intro}

The last years witnessed a renewed interest in configuration interaction (CI) approaches, and in particular in selected CI theories pioneered by Malrieu and co-workers\cite{Malrieu1973_CIPSI-Original,Malrieu1983_ImprovedCIPSI,Caballol1993_SelectedCI-EnergyDifference,Cabrero2002} that are, for instance, the foundation of the spectroscopy-oriented CI scheme by Neese.\cite{Neese2003_SORCI} Selected CI limits the cost of standard CI through an \textit{a priori} screening of the many-particle basis by evaluating \textit{a posteriori} the accuracy of this screening. Different flavors of selected CI are obtained by changing the criteria for these two steps. In the heath-bath CI (HBCI) theory introduced by Urmigar and co-workers,\cite{Holmes2016_HBCI,Sharma2017_HBCI} the screening is based on the magnitude of the CI matrix elements. Other options include comparisons with a reduced-size calculation, as in the projective CI of Evangelista\cite{Schriber2016_AdaptiveCI} and in the selected CI scheme by Head-Gordon and co-workers,\cite{HeadGordon2016_SelectedCI} or are based on an $n$-body expansion of the correlation energy, as investigated by Zimmermann and co-workers\cite{Ziemmermann2017_iCI,Zimmerman2019_Energy-Gradients} and by Gauss and co-workers.\cite{Eriksen2017_VirtualOrbitalSelectedCI,Gauss2018_MBE-FCI-WeaklyCorrelated} 

These developments need to be put into the context of two relatively new and highly efficient approaches to solve the full CI (or complete active space CI) problem. One is full CI Quantum Monte Carlo (FCIQMC) by Alavi and coworkers.\cite{Alavi2009_FCIQMCOriginal,Alavi2011_Initiator-FCIQMC} In FCIQMC, the diagonalization of the Hamiltonian is replaced by a stochastic sampling of the CI space through a Monte-Carlo algorithm in the electronic-configuration space. By contrast to early work of Greer,\cite{Greer1995_MonteCarlo-CI,Greer1998_MonteCarlo-CI} the CI coefficients are constructed from so-called walkers and the Fermion sign problem is avoided by walker annihilation.

The other one is the density matrix renormalization group (DMRG).\cite{White1992_DMRGBasis,White1993_DMRGBasis,Chan2008_Review,Zgid2009_Review,Schollwoeck2011_Review-DMRG,Chan2011,Wouters2014_Review,Keller2014,Kurashige2014_Review,Olivares2015_DMRGInPractice,Szalay2015_Review,Yanai2015,Knecht2016_Chimia} DMRG is an iterative optimization algorithm for wave functions parametrized in terms of so-called matrix product states (MPSs).\cite{Oestlund1995_MPS,Schollwoeck2011_Review-DMRG} Ground states of Hamiltonians featuring only short-range interactions can be represented by particularly compact MPSs,\cite{Hastings2007_AreaLaw} but this condition is rarely met for the full Coulomb Hamiltonian in electronic structure theory since each operator for the interaction of a pair of electrons couples four orbitals in its second-quantized form. Obviously, this situation does not at all resemble that of a nearest-neighbor interaction Hamiltonian, which would make DMRG iterations converge quickly. The number of variational parameters in a compact MPS scales only polynomially with system size, and therefore, the exponential scaling of full CI can be avoided for some target accuracy so that the curse of dimensionality is tamed. The DMRG energy is a nonlinear function of the tensors defining an MPS that are optimized iteratively during DMRG optimization. The advantages of non-linear expansions have already been exploited in other contexts, as in the multifacet graphically contracted CI by Shepard.\cite{Shepard2014_Mltifaced-I,Shepard2014_Multifaced-II} A major advantage of MPS over other parametrization schemes is the availability of DMRG as an efficient optimization scheme.

Early DMRG-CI applications to few-atom molecules\cite{Shuai1996_Original,Ciofini2000_DMRG,Fano2001_DMRG-QuantumChemistry,Chan2002_DMRG,Chan2003,Chan2004_EfficientDMRG,Legeza2003_DMRG-LiF,Legeza2003_DynamicalBlockState,Chan2004_NitrogenDMRG,Moritz2005_DKH-DMRG,Moritz2005_OrbitalOrdering,Marti2008_DMRGMetalComplexes} were soon followed by work on optimization of the orbitals\cite{Zgid2008_OrbitalOptimization,Ghosh2008_OrbitalOptimization,Yanai2009_DMRG-SCF,Legeza2016_OrbitalOptimization,Ma2017_QuadraticallyConvergent-DMRGSCF} and on perturbation theory.\cite{Kurashige2011_DMRGPT2,Morokuma2013_DMRG-PT2,Sharma2014_Hylleraas-DMRG,Kurashige2014_DMRG-CumulantExpansion,Roemelt2016_DMRGPT2,Chan2016_DMRG-NEVPT2,Sharma2016_QuasiDegenerate-PT,Freitag2017_DMRG-NEVPT2,Sharma2017_MRPT-DMRG,Sokolov2017_TDDMRG-Perturbation,Guo2018_StochasticPT-DMRG} Within only a decade, DMRG has been established as a reference method for electronic properties of large, strongly correlated systems. 

This perspective provides an overview of the application of DMRG to quantum chemical (QC) problems. Section~\ref{sec:DMRG_Theory} presents the main theoretical framework of DMRG, starting from its original formulation\cite{White1992_DMRGBasis} up to the most recent developments. Section~\ref{sec:DMRG+PT} discusses possible strategies to recover dynamical electron correlation. Sections~\ref{sec:vDMRG} and \ref{sec:TD-DMRG} present the extension of DMRG to vibrational and time-dependent (TD) problems, respectively. Section~\ref{sec:applications} highlights the most recent applications of DMRG to challenging strongly-correlated molecular systems.

\section{The density matrix renormalization group algorithm}
\label{sec:DMRG_Theory}

We first review the traditional presentation of DMRG with a focus on the optimization of ground states of the electronic Hamiltonian. Subsequent discussions then include energy-specific variants of DMRG, targeting of excited states, and multidimensional generalizations of DMRG.

\subsection{Elements of DMRG}

In 1992,\cite{White1992_DMRGBasis} DMRG was introduced by White as an improved version of Wilson's numerical renormalization group (NRG) approach.\cite{Wilson1075_NumericalRenormalization} Both, NRG and DMRG, approximate the ground state of an $N$-particle system based on the partitioning of the full quantum system into several blocks, each represented by at most $m$ basis functions (known as the renormalized basis). Blocks are then coupled together and iteratively optimized until convergence of a state for the complete system is reached. The block basis is truncated at each iteration step, keeping only $m$ elements to avoid the explosion of the number of basis states. The parameter $m$ is known as \quotes{bond dimension} or \quotes{number of renormalized block states}. It tunes both the accuracy and the computational demands of NRG and DMRG, which however differ in the criterion to truncate the basis. NRG keeps the $m$ lowest energy eigenfunctions of the Schr\"{o}dinger equation, while DMRG selects the $m$ lowest eigenfunctions of a reduced density matrix in order to produce a reduced-dimensional many-particle basis. This second choice has a more solid theoretical foundation since it provides the best approximation, in a least-squares sense, of the ground state wave function in terms of a linear combination of $m$ many-particle basis functions (each of which can be considered as iteratively refined contractions of determinants).\cite{White1993_DMRGBasis} This property explains the success of DMRG over NRG.

Since its first formulation, it has been clear that the efficiency of DMRG is maximal for one-dimensional systems. In this context, \quotes{one-dimensional} means that the one-particle states are sorted in such a way that they land on neighboring positions of a lattice resembling a short-range pair interaction. This sorting is known as the \quotes{DMRG lattice} and defines a linear iteration protocol. A formal proof of this property had been given 15 years after the introduction of DMRG, as a corollary of a theorem known as area law.\cite{Hastings2007_AreaLaw} The area law states that, for Hamiltonians containing only nearest-neighbour interactions and with a finite gap between the group and the first excited state, the entanglement entropy is constant in the limit of infinite size. A direct consequence is that the bond dimension $m$ needed to represent the ground state to a given accuracy becomes independent of the system size $L$. 

Intuitively, the area law requires that, if the quantum system is partitioned in two subsystems, the number of the interaction terms that couple the two subsystems in the Hamiltonian is independent of the total number of sites $L$. This implies that the entanglement between the two subsystems is also independent of the overall size $L$, and therefore, so is the bond dimension $m$. The success of DMRG for strictly one-dimensional spin systems is the reason why the first quantum chemical DMRG implementations were applied to the study of the $\pi$ electrons of conjugated polyenes, such as poly-para-phenylene. Their electronic properties were modeled with either the Hubbard\cite{Shuai1996_Original} or the Parisier-Parr-Pople\cite{Shuai1996_Original,Shuai1998_PPP,Shuai1998_Response-Polyene,Yaron1998_DMRG-ElectronHole,Fano1998_DMRG-PPP,Shuai1999_DMRGNonLinear,Bendazzoli1999_DMRG-PPP} Hamiltonian, the latter being a semiempirical Hamiltonian designed for $\pi$-conjugated systems. In both cases, only nearest-neighbor interactions are included and, therefore, the premises of the area law are met. Electronic properties are, however, governed by the full Coulomb Hamiltonian $\mathcal{H}_\text{el}$, which reads in second quantization

\begin{equation}
  \mathcal{H}_\text{el} = \sum_{pq}^{L'} h_{pq} \hat{a}_p^+ \hat{a}_q 
  + \frac{1}{2} \sum_{pqrs}^{L'} \langle pq || rs \rangle \hat{a}_p^+ \hat{a}_q^+ \hat{a}_s \hat{a}_r \, ,
  \label{eq:SQ_ElectronicStructure}
\end{equation} 
where $p$, $q$, $r$, and $s$ label different molecular orbitals and $h_{pq}$ and $\langle pq || rs \rangle$ are one- and two-electron integrals in the molecular orbital basis, respectively. The second term of Eq.~(\ref{eq:SQ_ElectronicStructure}) contains four-index integrals, whose range of interaction spans the molecular system. We highlight that the average interaction-range of a Hamiltonian, and therefore, the efficiency of DMRG, depends on the basis in which the Hamiltonian is expressed. For example, Legeza and co-workers showed this effect for the two-dimensional Fermi-Hubbard Hamiltonian that models interacting spins that are arranged on a square lattice. The magnitude of  the long-range interactions is reduced if a momentum space-representation is adopted that leads to an increase of efficiency of DMRG. Similarly, orbital localization can lead to a more compact representation of the Coulomb interaction. The presence of these long-range interactions made the first applications of DMRG to quantum chemistry not as efficient as for the model Hamiltonians in solid physics. The first DMRG implementation for the quantum chemical Hamiltonian of Eq.~(\ref{eq:SQ_ElectronicStructure}) was presented in 1999 by White and Martin.\cite{White1999} This work was followed by a rapid development of quantum-chemical applications of DMRG owing to the work of several groups, including Mitrushenkov et al.,\cite{Fano2001_DMRG-QuantumChemistry,Fano2003_DMRG-Second} Daul et al.,\cite{Ciofini2000_DMRG} Chan and co-workers,\cite{Chan2002_DMRG,Chan2003,Chan2004_EfficientDMRG,Chan2004_NitrogenDMRG,Chan2005_DMRG-NonOrthogonal} and Legeza, Hess, and co-workers.\cite{Legeza2003_DMRG-LiF,Legeza2003_DynamicalBlockState,Legeza2003_OrderingOptimization,Legeza2004_DataCompression-DMRG,Moritz2005_OrbitalOrdering} Naturally, these pilot applications focused on full-CI energies of small molecules with up to six atoms. Later studies applied DMRG as a CAS-CI solver for active spaces with up to 100 orbitals and they extended its range of applicability to large molecules. Since the non-relativistic Schr\"{o}dinger equation does not meet the conditions of the area law, the bond dimension $m$ required to obtain converged energies with a given accuracy will depend on the lattice size $L$, \textit{i.e.} on the number of orbitals. Nevertheless, these first quantum-chemical applications of DMRG showed that $m$ is largely independent of $L$ and, therefore, DMRG renders full- and CAS-CI calculations on systems with up to 100 orbitals feasible. In particular, DMRG turned out to be efficient even for compact non-linear systems such as transition metal complexes and clusters\cite{Marti2008_DMRGMetalComplexes}.

We have already mentioned that the efficiency of DMRG is due to the fast convergence of the energy with respect to the bond dimension $m$. Nevertheless, a full-CI wave function is strictly equivalent to an MPS with a bond dimension $m$ that grows exponentially with $L$. It is therefore natural to increase the efficiency of DMRG based on the same strategies that have already been developed for truncated CI calculations. For example, the molecular orbitals can be optimized together with the state coefficients, as in complete active space self-consistent field (CAS-SCF) approaches. An efficient strategy to couple SCF and DMRG, usually known as DMRG-SCF, was introduced by Zgid and Nooijen\cite{Zgid2008_OrbitalOptimization} and others.\cite{Ghosh2008_OrbitalOptimization,Yanai2009_DMRG-SCF,Ma2017_QuadraticallyConvergent-DMRGSCF,Chan2017_DMRG-SCF-SecondOrder}

A further increase in efficiency is achieved by exploiting the symmetry of the Hamiltonian. In standard CAS calculations, symmetry constraints induce a block structure of the full Hamiltonian matrix in the CI basis. Even if in DMRG this matrix is never calculated explicitly, its local representations built at each DMRG iteration will also have a block structure.\cite{McCulloch2007_FromMPStoDMRG} This property can be exploited to reduce the computational demands of the optimization. For Abelian groups, such as U(1) describing the conservation of the particles number, the development of a symmetry-adapted DMRG algorithm requires only minor modifications to the standard implementation.\cite{Vidal2011_DMRG-U1Symm,Troyer2011_PEPS-Symmetry} The extension to spin symmetries is less trivial because they are described in terms of a non-Abelian group, namely SU(2). The first attempt to derive a spin-adapted formulation of DMRG was proposed by Zgid and Noojien in 2008.\cite{McCulloch2002_NonAbelianDMRG,Zgid2008_DMRGSpinAdaptation} However, this approach allowed only to enforce the spin symmetry \textit{a posteriori}, at the end of each DMRG microiteration. An SU(2) invariant formulation of DMRG that exploits spin symmetry to reduce the number of parameters of the MPS was introduced later by Sharma and Chan\cite{Sharma2012_SpinAdapted-DMRG,Sharma2015_GeneralNonAbelian} and by Wouters and co-workers.\cite{Wouters2012_SpinAdapted} As this symmetry is difficult to implement, it has been argued\cite{Chan2017_SpinAdapted-DMRG} that a broken-symmetry wave function optimization with a subsequent spin projection can be very efficient, as considered also in traditional approaches.\cite{Scuseria2011_ProjectedQuasiparticle,Scuseria2012_ProjectedHF,Ten-no2016_ProjectedCI}

The efficiency of DMRG can also be increased by tuning DMRG-specific parameters. For example, sites (\textit{e.g.} orbitals, in the electronic-structure case) can be sorted on the DMRG lattice to place strongly entangled ones close to one another to reduce long-range correlations. An optimized ordering can be obtained either from interaction measures derived from one- and two-electron integrals,\cite{Chan2002_DMRG} with genetic algorithms,\cite{Moritz2005_OrbitalOrdering} or, very successfully, through a Fiedler vector ordering based on entanglement orbital entropies,\cite{Legeza2003_OrderingOptimization,Rissler2006_QuantumInformationOrbitals} whose definition will be discussed in more detail below. Converged orbital entropies can be obtained from partially converged DMRG results, carried out with a low value of $m$. The resulting optimized sorting can be employed in more efficient DMRG calculations. 

Canonical HF orbitals can be strongly delocalized, enhancing long-range interactions. With localized orbitals obtained by a unitary transformation of the HF orbitals long-range interactions can be minimized, increasing the efficiency of DMRG.\cite{Fano2012_OrbitalLocalization,Wouters2015_CheMPS2}

A proper inclusion of relativistic effects requires the generalization of MPSs to the symmetries of the Dirac Hamiltonian.\cite{Moritz2005_DKH-DMRG,Knecht2014} Its symmetry properties will not be described in terms of the SU(2) group if the Hamiltonian includes spin-orbit coupling operators, but in terms of double groups, coupling spatial and spin symmetry.\cite{Battaglia2018_RelativisticDMRG}

\subsection{MPS/MPO formulation of DMRG}

A main limitation of the original formulation of DMRG\cite{White1992_DMRGBasis,White1993_DMRGBasis} is the lack of a specific \textit{ansatz} for the wave function $| \Phi \rangle$. However, shortly after its introduction,\cite{Oestlund1995_MPS,Rommer1997_MPS-Ansatz} it was shown that DMRG iteratively builds a wave function that can be expressed as,

\begin{equation}
 \begin{aligned}
  | \Phi \rangle = \sum_{\boldsymbol{\sigma}} \sum_{a_1=1}^{m} \cdots \sum_{a_{L-1}=1}^{m} &
    M_{1,a_i}^{\sigma_1} M_{a_1,a_2}^{\sigma_2} \cdots M_{a_{L-1},1}^{\sigma_L} \\
    \times & | \sigma_1 \sigma_2 \cdots \sigma_L \rangle \, ,
 \end{aligned}
 \label{eq:MPS}
\end{equation}
where $L$ is the number of sites and $| \sigma_1 \sigma_2 \ldots \sigma_L \rangle$ are occupation number vectors and is equivalent to the number of orbitals $L'$ included in the Hamiltonian of Eq.~(\ref{eq:SQ_ElectronicStructure}). In full CI, $L$ is equal to the basis set size, while for CAS-CI it is the number of orbitals in the CAS. $M_{a_i,a_{i+1}}^{\sigma_{i+1}}$ are three-dimensional tensors with dimensions $N_{i+1} \times m \times m$, where $N_{i+1}$ is the dimension of the local basis at the ($i$+$1$)-th site. The parametrization of Eq.~(\ref{eq:MPS}) defines an MPS. By analogy with Eq.~(\ref{eq:MPS}), operators can also be expressed in a corresponding format reflecting the site structure of the DMRG lattice,\cite{McCulloch2007_FromMPStoDMRG}

\begin{equation}
 \begin{aligned}
  \mathcal{W} = \sum_{\boldsymbol{\sigma},\boldsymbol{\sigma}'} 
  \sum_{b_1=1}^{r_1'} \cdots \sum_{b_L=1}^{r_L'}
    & W_{1,b_i}^{\sigma_1,\sigma_1'} W_{b_1,b_2}^{\sigma_2,\sigma_2'} \cdots 
  W_{b_{L-1},1}^{\sigma_L,\sigma_L'} \\
    & \times | \sigma_1 \sigma_2 \cdots \sigma_L \rangle \langle \sigma_1' \sigma_2' \cdots \sigma_L' | \, ,
 \end{aligned}
 \label{eq:MPO}
\end{equation}
known as the Matrix Product Operator (MPO) format. Unlike Eq.~(\ref{eq:MPS}), which represents an approximation of a wave function, whose accuracy depends on $m$, Eq.~(\ref{eq:MPO}) is exact and the $r_i'$ parameters depend on the specific form of the operator. The $r_i'$ grow with the maximum length of second-quantized operator strings appearing in the definition of $\mathcal{W}$. Different algorithms to construct MPO representations of operators starting from their second-quantization form have been proposed,\cite{Frowis2010_MPOGeneric,Dolfi2014_ALPSProject,Hubig2017_GenericMPO} some of which are general enough to be applied to the quantum chemical Hamiltonians.\cite{Keller2015_MPSMPODMRG,Chan2016_MPO-MPS} Eqs.~(\ref{eq:MPS}) and (\ref{eq:MPO}) can be combined to determine the energy expectation value $E\left[| \Phi \rangle \right]$. Minimization of $E\left[| \Phi \rangle \right]$ yields the best approximation of the ground-state wave function as an MPS in a variational sense. This minimization is carried out with respect to variations of the entries of the tensor for site $i$ ($M_{a_{i-1}a_i}^{\sigma_i}$), while keeping all the other ones fixed. Iterating this minimization along the lattice leads to a DMRG sweep. Instead of optimizing a single tensor per micro-iteration, in the so-called two-sites optimization two consecutive tensors are optimized simultaneously. In practice, the energy is minimized with respect to the entries of the two-site tensor $T_{a_{i-1},a_{i+1}}^{\sigma_i,\sigma_{i+1}}$, defined as

\begin{equation}
  T_{a_{i-1},a_{i+1}}^{\sigma_i,\sigma_{i+1}} 
    = \sum_{a_i=1}^m M_{a_{i-1},a_i}^{\sigma_i} M_{a_i,a_{i+1}}^{\sigma_{i+1}} \, .
  \label{eq:TwoSiteTensor}
\end{equation}

After optimization, the single-site tensors ($M_{a_{i-1},a_i}^{\sigma_i}$ and $M_{a_i,a_{i+1}}^{\sigma_{i+1}}$) are recovered from the singular value decomposition (SVD) of $T_{a_{i-1},a_{i+1}}^{\sigma_i,\sigma_{i+1}}$. However, the rank of the two-site tensor after optimization may be larger than the one of the original tensors ($m$ in Eq.~(\ref{eq:TwoSiteTensor})) and, therefore, the SVD must be truncated to keep the bond dimension fixed. Alternatively, the bond dimension $m$ can be adapted in order to keep the truncation error fixed. This second alternative, that is employed in the so-called dynamical block state selection (DBSS) scheme,\cite{Legeza2003_DynamicalBlockState} enables one to adapt the bond dimension dynamically, based on a target accuracy for the wave function.

This alternative formulation of DMRG, usually referred to as MPS/MPO (or second-generation) formalism, is formally equivalent to the original DMRG theory.\cite{Schollwoeck2011_Review-DMRG,Chan2016_MPO-MPS} Nevertheless, the MPS/MPO formulation is a more flexible framework to apply DMRG to complex Hamiltonians, especially when containing long-range interactions. In fact, the original, first-generation formulation of DMRG constructs the representation of the Hamiltonian from the one of each elementary second-quantization operator in the system/environment basis that is set up in each microiteration step. This construction becomes quickly cumbersome for operators represented by long strings of elementary operators, such as $\mathcal{H}^2$. From $\mathcal{H}^2$, the energy variance for a given state can be obtained,\cite{Filippi2005_Variance} which is a reliable metric to assess the accuracy of DMRG.\cite{Hubig2018_ErrorEstimates} Similarly, we derived an MPS/MPO version\cite{Keller2016_SpinAdapted} of the the SU(2)-invariant formulation of DMRG originally derived in a first-generation framework. In the MPS/MPO formulation the complexity of a single microiteration step is independent of the form of the operator, provided that it can be encoded as an MPO. The availability of a general algorithm for constructing MPO representations of operators of arbitrary complexity therefore makes the MPS/MPO framework much easier to extend beyond the calculation of ground state energies.

MPSs have been considered also from a numerical analysis point of view and are known in that context by the name of tensor train (TT) factorization.\cite{Oseledets2009_TuckerFormat,Oseledets2011_TTGeneral} The TT theory is not limited to the solution of the Schr\"{o}dinger equation, but can be applied to solve a wider range of equations.\cite{Dolgov2012_TT-FokkerPlanck,Kazeev2014_TT-MasterEquation} Some of the algorithms already known for DMRG have been later generalized to TT theory. For instance, the sweep-based DMRG optimization is known as alternating least squares (ALS) in the TT context.\cite{Holz2012_ALSTheory} Conversely, other algorithms, originally devised for TTs, have been later extended to DMRG. This is the case for the calculation of multiple eigenpairs of an operator with ALS,\cite{Dolgov2014_BlockEigenvalues} which has been applied to optimize excited states with DMRG.\cite{Verstraete2012_vNRG} 

\subsection{Targeting excited states with DMRG}

The area law, which provides a theoretical foundation of DMRG, implies that ground states of Hamiltonians with short-range interactions and with a non-zero gap between the ground and the first excited state can be represented as MPSs with a bond dimension $m$ that is independent of system size. Owing to the generalizations of ALS to the simultaneous optimization of multiple eigenpairs, excited states can be targeted with DMRG. However, the reliability of representing excited states with compact MPSs is not guaranteed, and, hence, neither is the fast convergence of DMRG with respect to the bond dimension $m$. Recently, the area law has been generalized to states which can be encoded as many-body localized states,\cite{Bauer2013_AreaLaw-ManyBodyLocalized,Serbyn2013_AreaLaw-MBL} \textit{i.e.} states which are localized on a small portion of the DMRG lattice, and therefore, can be described in terms of excitations involving only a small subset of the $L$ sites composing the full system. Several, non-equivalent definitions of many-body localized states have been derived. For some of them,\cite{Bauer2013_AreaLaw-ManyBodyLocalized} it was shown that they can be encoded as compact MPSs. For some model Hamiltonians, it has even been postulated that any eigenstate is a many-body localized state.\cite{Oganesyan2007_ManyBodyLocalization} These generalizations of the area law has promoted the design of excited-state variants of ground-state DMRG. 

In first-generation DMRG, the ground state is iteratively approximated as a linear combination of the eigenfunctions of local density matrices. They are in turn obtained from the approximated wave function calculated in the previous iterations. Excited states can be approximated by tracking higher-energy eigenstates of the same local density matrices, but the basis in which these states are represented is optimal, in a least-squares sense, only for the ground state. Therefore, its accuracy may deteriorate when applied to excited states. This problem can be alleviated by exploiting state-average density matrices for the construction of the renormalized basis in order to produce a balanced representation of all relevant states.\cite{Dorando2007_TargetingExcitedStates,Hu2015_DMRGExcitedStateOptimization} However, the state-average density matrix is not optimal for any state and this slows down the convergence rate of DMRG with respect to $m$ and renders such state-average approaches unpractical when a large number of excited states is targeted.

Moreover, in the MPS/MPO framework, the availability of a well-defined energy functional, whose minimization provides the DMRG wave function, renders the extension to excited-state targeting simpler. Excited states can be, for instance, optimized sequentially with state-specific algorithms. After the optimization of the ground state $| \Phi_0 \rangle$, the first excited state is obtained from a constrained, variational optimization in the space orthogonal to the ground state.\cite{McCulloch2007_FromMPStoDMRG,Keller2015_MPSMPODMRG} This is achieved by replacing the Hamiltonian $\mathcal{H}$ with its projected counterpart $\mathcal{H}^p$, defined as

\begin{equation}
  \mathcal{H}^p = \left( \mathcal{I} - | \Phi_0 \rangle \langle \Phi_0 | \right)
  \mathcal{H}
  \left( \mathcal{I} - | \Phi_0 \rangle \langle \Phi_0 | \right) \, .
  \label{eq:H_projected}
\end{equation}

All terms appearing in Eq.~(\ref{eq:H_projected}) can be encoded as MPOs,\cite{Schollwoeck2011_Review-DMRG,Keller2015_MPSMPODMRG} and therefore, the ground state of $\mathcal{H}^p$ (\textit{i.e.} the first excited state of $\mathcal{H}$) can be optimized with the standard DMRG algorithm. Higher-lying excited states are then obtained from successive constrained optimizations.

The need to optimize the states in increasing order of the energy restricts the algorithms introduced above to the optimization of low-lying excited states. However, several applications require the calculation of high-energy eigenstates. In electronic structure calculations, the simulation of X-ray spectra involves high-energy electronically excited states.\cite{Norman2018} The same requirement holds for vibrational structure calculations in the fingerprint region (in the energy range 800-2000~cm$^{-1}$). The optimization of such high-energy states may not be trivial if the diagonalization is replaced by the minimization of the energy functional with respect to the MPS entries and if the global minimum is associated to the ground state only. This problem can be circumvented by mapping the Hamiltonian onto an auxiliary operator, whose ground state is one of the excited states of the original Hamiltonian.\cite{Leforestier1993_SpectralTransformation,Ventra2002_TargetedDiagonalization,Zuev2015_GeneralEigensolver} 

For example, the ground state of the shift-and-invert (S\&I) operator $\Omega_\omega$,

\begin{equation}
  \Omega_\omega = \left( \omega - \mathcal{H} \right)^{-1}
  \label{eq:SandI_definition}
\end{equation}
is that lowest excited state of $\mathcal{H}$ with an energy larger than $\omega$. Hence, a DMRG optimization, if applied to $\Omega_\omega$, will approximate the excited state with energy closest to $\omega$. This approach, denoted DMRG[S\&I],\cite{Yu2017_ShiftAndInvertMPS,Baiardi2019_HighEnergy-vDMRG} has two limitations. First, the choice of the shift parameter $\omega$ requires an estimate of the energy of the target state. The accuracy of this estimate must be high in regions with a high density of states, in which small variations of $\omega$ can lead to convergence of undesired states. This limitation can be lifted by combining the S\&I scheme with a maximum-overlap (MaxO) criterion\cite{Kammer1976_RootHoming,Reiher2004_ModeTracking,Gill2009_SCF-ExcitedStates} by which the state with the largest overlap with a predetermined MPS is followed. The maximum-overlap criterion improves significantly the stability of DMRG[S\&I] since states close in energy are often localized on different parts of the DMRG lattice. The predetermined MPS can be chosen, for instance, from the eigenstates of the non-interacting part of the Hamiltonian. MaxO-based formulations of DMRG have been recently applied to the Hubbard Hamiltonian,\cite{Khemani2016_ExcitedStateDMRGSpatial,Devakul2017} under the name DMRG-XX, and to vibrational Hamiltonians.\cite{Baiardi2019_HighEnergy-vDMRG}

\begin{figure}
  \centering
  \includegraphics[width=.35\textwidth]{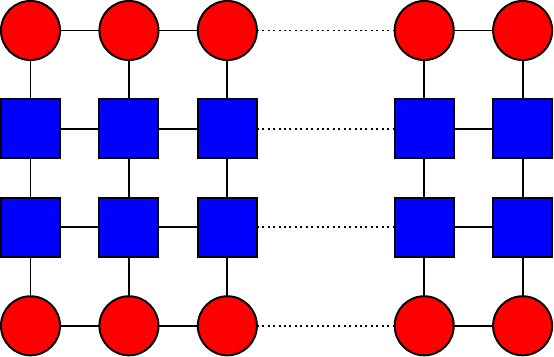}
  \caption{Tensor network associated to the evaluation of the expectation value of $\mathcal{H}^2$ over an MPS $| \Phi \rangle$. Red circles are associated to the entries of the MPS ($M_{a_{i-1,a_i}}^{\sigma_i}$), blue squares are associated to the entries of the MPO ($W_{b_{i-1},b_i}^{\sigma_i,\sigma_i'}$).}
  \label{fig:H_Squared}
\end{figure}

Another issue associated with DMRG[S\&I] is the selection of the operator $\mathcal{H}$ for Eq.~(\ref{eq:SandI_definition}). Choosing $\mathcal{H}$ to be the full Hamiltonian of the system would require explicit inversion of MPOs. It has recently been demonstrated\cite{Yu2017_ShiftAndInvertMPS,Villalonga2018_SI-MPS} that the inversion can be avoided, but the resulting equations involve expectation values of the squared Hamiltonian $\mathcal{H}^2$. As discussed in Ref.~\citenum{Schollwoeck2011_Review-DMRG}, the exact evaluation of matrix elements of $\mathcal{H}^2$ is simplified in an MPS/MPO-based DMRG implementation. The tensor network that must be contracted to calculate the expectation value $\mathcal{H}^2$ over an MPS is given in Figure~\ref{fig:H_Squared}. However, it is obviously computationally more expensive than for the standard Hamiltonian $\mathcal{H}$. Analogous equations are obtained with the folded operator $\Omega_\omega^F$,

\begin{equation}
  \Omega_\omega^F = \left( \omega - \mathcal{H} \right)^2 \, ,
  \label{eq:Folded_definition}
\end{equation}
as auxiliary operator. We recently employed this folded operator to target excited states with DMRG for vibrational problems.\cite{Baiardi2019_HighEnergy-vDMRG} The computational cost associated with the evaluation of Eq.~(\ref{eq:Folded_definition}) can be reduced if the auxiliary operators are obtained from the local representation of the full Hamiltonian in the renormalized basis which is constructed at each DMRG microiteration step.\cite{Dorando2007_TargetingExcitedStates,Lim2016_EnergySpecific-DMRG} However, as noted in Ref.~\citenum{Devakul2017}, convergence is not guaranteed within the latter schemes, since the resulting equations do not correspond to the minimization of any energy functional.

\subsection{Multidimensional generalizations}

The main feature of the MPS parametrization of Eq.~(\ref{eq:MPS}) is that only matrices centered on neighboring sites are contracted together. We have already mentioned above that this contraction pattern is designed to describe efficiently one-dimensional quantum systems, represented by Hamiltonians in which the entanglement between two sites decays with their distance on the DMRG lattice. The decay rate of the entanglement is, however, determined by the average length scale of the interactions in the Hamiltonian. For this reason, the convergence of DMRG iterations will be slower and higher values of $m$ will be needed to obtain converged energies if applied to ground states of general Hamiltonians containing long-range interactions.

To increase the efficiency of DMRG for more complex Hamiltonians, the MPS parametrization can be generalized to wave functions known as tensor network states (TNSs). Similar to MPSs, a tensor is associated with each site of a lattice, but of arbitrary shape. Accordingly, these tensors can have more than two auxiliary indices, (cf. the $a_i$ in Eq.~(\ref{eq:MPS})) and are contracted together following more complex patterns. For this reason, TNSs are usually viewed as multidimensional generalizations of MPSs.

Despite the successful application of TNS in physics,\cite{Isacsson2006_PEPS,Verstraete2007_PEPS-HardCoreBosons,Cirac2008_PEPS,Chan2009_CorrelatorMPS,Wei2011_AKLT-2D,Orus2014_GuidePEPS} their extension to quantum chemical problems has been limited by two issues. The success of DMRG relies on the availability of ALS, which reduces a complex non-linear optimization problem to a series of standard eigenvalue problems. Generalizations of ALS to arbitrary forms of TNSs are, however, currently not known. For this reason, tensor network states are usually optimized with a stochastic Monte Carlo evaluation of the energy integral,\cite{Vidal2007_QMC-TensorNetwork,Verstraete2008_StringState-QMC,Marti2010_CGTN,Vidal2012_PerfectSampling} although this is much less efficient than ALS. Among the TNS parametrizations proposed in the literature, those which have been most successfully applied to quantum chemical problems are built from so-called tree tensor networks states (TTNS)\cite{Legeza2010_TreeTensorNetworks,Legeza2018_T3NS} which exploit an iterative optimization scheme that resembles ALS. TTNSs map the orbitals to a tree-structured lattice in which groups of $n$ sites (the parameter $n$ is known as order of the TTNS) are first correlated together. The resulting renormalized bases are correlated again in groups of $n$ elements until all sites are included. TTNSs can be interpreted as a hierarchical generalization of a MPS where $n$ orbitals are correlated together with standard DMRG and the resulting MPSs are employed as a local basis for another MPS. The active space decomposition (ASD) algorithm introduced by Shiozaki and co-workers\cite{Shiozaki2013_ASD-General,Shiozaki2015_ASD-OrbitalOptimization} to calculate multi-configurational wave function for molecular aggregates relies on a similar parametrization. ASD expresses the wave function for the aggregate $| \Psi_\text{ASD} \rangle$ as

\begin{equation}
  | \Psi_\text{ASD} \rangle = \sum_{i_1,\ldots,i_{n_M}} C_{i_1,\ldots,i_{n_M}} | i_1 \cdots i_{n_M} \rangle \, ,
  \label{eq:ASD_Expansion}
\end{equation} 
where $n_M$ is the number of monomers in the aggregate and $| i_j \rangle $ is a complete basis for the $j$-th monomer that is obtained from CAS-SCF. Eq.~(\ref{eq:ASD_Expansion}) is a full-CI expansion that suffers, for large aggregates, from the problem of the curse of the dimensionality. Parker and Shiozaki suggested to tame the high computational cost with DMRG\cite{Shiozaki2014_ASD-DMRG} by replacing Eq.~(\ref{eq:ASD_Expansion}) with a MPS, where the lattice size is equal to the number of monomers of the aggregate, and the local basis is the CAS-SCF basis of a single monomer. If the local basis is obtained from DMRG instead of from standard CAS-SCF, the resulting wave function would be an example of hierarchical DMRG treatment. Such an approach has been exploited by Nishio and Kurashige to calculate correlated wave functions of molecular aggregates.\cite{Kurashige2019_LowRank-ASD} The ground and low-energy excited states of each monomer are encoded as MPSs and optimized with DMRG. The wave function of the aggregate is then expressed as in Eq.~(\ref{eq:ASD_Expansion}) from the resulting basis of MPSs. Unlike ASD-DMRG, which approximates $C_{i_1,\ldots,i_{n_M}}$ as an MPS, in Ref.~\citenum{Kurashige2019_LowRank-ASD} the tensor is replaced by its rank-one factorization. Such approximation reduces significantly the number of variational parameter, but does not encode efficiently strong entanglement between the monomers. For this reason, the rank-one factorization is particularly efficient for weakly bonded molecular aggregates.\cite{Kurashige2019_LowRank-ASD} Conversely, the ASD-DMRG scheme is expected to be more effective in presence of strong entanglement, such as for chemically bonded monomers.

The localized active space SCF (LAS-SCF) approach introduced by Hermes and Gagliardi\cite{Hermes2019_LAS-SCF} is another example of a multi-layer CAS-SCF scheme. LAS-SCF expresses the wave function of an aggregate as

\begin{equation}
  | \Psi_\text{LAS} \rangle = \prod_{i=1}^{n_M} \psi_\text{CAS}^{(i)} \, ,
  \label{eq:LAS-SCF}
\end{equation}
where $\psi_\text{CAS}^{(i)}$ is a CAS-SCF wave function for the $i$-th fragment. The entanglement between each monomer, that is included in ASD through a full-CI expansion, might seem absent in the LAS-SCF wave function of Eq.~(\ref{eq:LAS-SCF}). However, the interaction between monomers is included during the orbital optimization with the density matrix embedding theory (DMET).\cite{Chan2012_DMET,Chan2013_DMET-StrongCoupling,Chan2016_DMET-Review} 

Both the ASD and LAS-SCF wave functions can be applied to any molecular system if the orbitals are partitioned in $n_M$ groups. However, even if the choice for this partition is trivial for molecular aggregates, it is not straightforward for more general molecules in which the orbitals are not localized on a different portion of the molecules. This choice could be automatized, for example, based on quantum-information measures, such as the two-orbital entropy\cite{Legeza2003_OrderingOptimization,Rissler2006_QuantumInformationOrbitals,Boguslawski2012_OrbitalEntanglement} obtained from a partially converged DMRG calculation following the same idea as for the \texttt{AutoCAS} algorithm that will be presented in Section~\ref{sec:applications}.\cite{Stein2016_AutomatedSelection}

An additional limitation which has impeded a widespread application of TNSs to quantum chemical problems has been the lack of a parametrization providing an adequate compromise between flexibility and computational cost of the optimization. TNSs designed for regular interaction patterns are not general enough to be applied to quantum chemical problems. This holds true for projected entangled pair states,\cite{Cirac2008_PEPS} designed to describe two-dimensional spin lattices, which are not appropriate when applied to Hamiltonians without such specific, regular interaction patterns.\cite{Legeza2010_TreeTensorNetworks} Moreover, the computational cost associated with the optimization of more general TNS parametrizations quickly becomes intractable. This has been observed, for example, for complete graph tensor network states.\cite{Chan2009_CorrelatorMPS,Marti2010_CGTN,Kovyrshin2017_SelfAdaptiveTN} Such general parametrization can reproduce, in principle, strong entanglement between any set of orbitals but the price to pay is a steep increase of the variational parameters that then needs to be tamed by sequential optimization schemes.\cite{Kovyrshin2017_SelfAdaptiveTN}

\section{A major challenge: recovering dynamical correlation}
\label{sec:DMRG+PT}

Owing to the limited number of basis states of the lattice, DMRG is usually applied as a CAS approach. As any CAS-based approach, it efficiently recovers static correlation, \textit{i.e.} the portion of electron correlation connected to the occurrence of more than one dominant Slater determinant in the CI wave function. For the inclusion of dynamical correlation from those basis states omitted from the CAS, DMRG must be coupled to approaches that can capture these contributions such as perturbation theories, coupled-cluster-based methods, and short-range density functional theory as will be discussed in the following.

\subsection{MPS-based perturbation theories}

Perturbation theory represents the most common way of assessing dynamical correlation effects. Perturbation approaches differ in the choice of the reference Hamiltonian and can be derived to different orders. CAS perturbation theory to the second order (CASPT2)\cite{Roos1990_CASPT2,Roos1992_CASPT2} starts from the generalized Fock operator as the reference Hamiltonian. The bottleneck of CASPT2 calculations, as for any other multireference perturbative approach, is the evaluation of three- and four-body density matrix elements. The MPS parametrization, together with a cumulant expansion of the density matrices, has been exploited to approximate these high-order reduced density matrices\cite{Kurashige2014_DMRG-CumulantExpansion} and to reduce the computational effort of the perturbation step with respect to standard CASPT2. In addition to such approximations, DMRG-PT2\cite{Kurashige2011_DMRGPT2,Morokuma2013_DMRG-PT2} suffers, as any perturbation theory, from numerical instabilities in the presence of nearly degenerate states (also known as intruder states). These instabilities can be avoided by introducing level shifts in the reference Hamiltonian to artificially increase the energy of intruder states.\cite{Roos1996_LevelShift,Malmqvist1997_ImaginaryLevelShift} A more elegant alternative, not depending on any external shift parameter, is to change the zeroth-order Hamiltonian. A reliable choice has been demonstrated to be the Dyall Hamiltonian,\cite{Dyall1995_Hamiltonian-NEVPT2} which includes, in addition to the standard CAS contributions, the M{\o}ller-Plesset reference Hamiltonian for the core and virtual orbitals. Perturbation theory relying on the Dyall Hamiltonian is called $n$-electron valence second-order perturbation theory (NEVPT2)\cite{Angeli2002_NEVPT2} which is more stable than standard CASPT2. Also NEVPT2 has been built on top of DMRG wave functions.\cite{Roemelt2016_DMRGPT2,Knecht2016_Chimia,Sokolov2017_TDDMRG-Perturbation,Freitag2017_DMRG-NEVPT2,Sharma2017_MRPT-DMRG}

Other perturbative approaches have been introduced that do not require the calculation of high-order density matrices. Coupling them with DMRG could enable one to target larger active spaces.
The driven similarity renormalization group (DSRG) by Evangelista\cite{Evangelista2014_DSRG} replaces the diagonalization of the CAS Hamiltonian by a sequence of unitary transformations which progressively decouple the basis starting from the elements with a higher energy separation. This algorithm will be equivalent to full-CI if the sequence of transformations is driven to convergence. Conversely, if it is stopped at an intermediate decoupling degree, only determinants with significant energy difference will be decoupled and nearly-degenerate states (\textit{i.e.}, intruder states) will be left unchanged. The resulting basis is a reliable reference for perturbation theory not suffering from instabilities.\cite{Evangelista2017_DSRG-PT3} When applied to multideterminant wave functions, DSRG represents a cost-effective alternative to CASPT2 and NEVPT2, since only three-body reduced density matrices are required. However, although the multireference generalization of DSRG is known,\cite{Li2015_DSRG-PT2} together with its coupling with second-order perturbation theory,\cite{Hannon2016_DSRG-Cholesky} its further extension to MPS wave functions has not been explored yet. The accuracy of DSRG depends, in principle, on the degree of coupling at which the block diagonalization is stopped. DSRG parametrizes the unitary transformation through a flow parameter $s$ whose inverse is related to the maximum energy difference for which states are decoupled. This would suggest that the value of the flow parameter at which the transformation is stopped, $s_\text{max}$, is the DSRG equivalent of the imaginary shift of CASPT2. Pilot studies indicated,\cite{Evangelista2017_DSRG-PT3} however, that DSRG is significantly more stable than CASPT2 upon changes of $s_\text{max}$.

The random phase approximation (RPA)-based theory introduced by Pernal\cite{Pernal2018_ExtendedRPACorrelation} represents another promising cost-effective perturbative scheme, in which dynamical correlation is obtained from an adiabatic connection formula. The correlation energy is expressed as an integral of quantities depending on one- and two-body reduced density matrices only, obtained through adiabatically switching on the correlation potential. In the original work,\cite{Pernal2018_ExtendedRPACorrelation} two-body reduced density matrices were expressed in terms of the one-electron transition matrices, which in turn were obtained from extended RPA equations.\cite{Pernal2012_ExtendedRPA} This RPA-based theory has been generalized to multi-reference wave functions\cite{Pastorczak2018_ExtendedRPA-CAS} under the assumption that the occupation of the CAS orbitals is approximately constant during the adiabatic switch-on of the electron-electron interaction. This assumption will be valid only if the CAS is big enough to include all static correlation effects. In this respect, the coupling of this RPA-based approach with DMRG is particularly appealing. This would require the extension of RPA to wave functions expressed as MPSs and could be accomplished within the recently introduced time-dependent formulation of DMRG.\cite{Dorando2009_AnalyticalResponseFunction,Nakatani2014_LinearResponseDMRG}

Any perturbation theory can be efficiently coupled to DMRG if the structure of MPSs can be exploited to speed-up the evaluation of the perturbative correction. As we have already mentioned above, this is not the case for CASPT2 or NEVPT2 that are based on sum-over-states expression and require the calculation of high-order reduced density matrices. The first-order correction to a wave function can be calculated as the minimum of the so-called Hylleraas functional.\cite{Sinanoglu1961_Hylleraas} Second-order correction to the energy can then be obtained trivially from the well-known ($2n$+$1$) rule. The reformulation of perturbation theory as a variational problem is particularly appealing in connection with DMRG because it allows one to derive perturbative corrections by applying ALS as for ground-state optimization. This idea, introduced by Chan and Sharma,\cite{Sharma2014_Hylleraas-DMRG} has recently been applied to quasi-degenerate\cite{Sharma2016_QuasiDegenerate-PT} and multireference perturbation theory.\cite{Sharma2017_MRPT-DMRG} Such Hylleras-based perturbative scheme will be, however, efficient, only if the first-order correction of the wave function can be represented as an MPS with a low bond dimension $m$ and, as has been discussed by Chan and co-workers,\cite{Guo2018_DMRG-Hylleraas} this is not the case for large active spaces. To reduce the size of the first-order correction MPS, it has been first proposed to express it as a sum of MPS, each with a smaller value of $m$.\cite{Guo2018_DMRG-Hylleraas} In alternative, the perturbative correction can be expressed as an average over the wave function probability density, as done for selected CI,\cite{Sharma2017_HBCI,Loos2017_StochasticPT} and evaluated stochastically. This second scheme is particularly appealing thanks to the availability of algorithm to sample efficiently configurations from the probability distribution of a MPS.\cite{White2009_MinimallyEntangled,Vidal2012_PerfectSampling}

\subsection{Combining the MPS with coupled cluster parametrizations}

Coupled cluster (CC) is the reference method to study electronic properties of single-reference systems lacking strong static correlation. For this reason, several recent attempts to apply CC corrections to multi-determinant wave functions\cite{Jeziorski2010_MRCC,Musial2011_MRCC-Review,Kohn2013_SS-MRCC-Review,Evangelista2018_MRCC-Review} have a natural extension to DMRG.

Different multi-reference generalizations of CC have been proposed. They may be classified, in broad terms, as internally-contracted (ic) multireference CC (MRCC) and Jeziorski-Monkhorst CC (details about the theory can be found, for example, in a recent review\cite{Evangelista2018_MRCC-Review}). The former formulation applies a unique CC exponential operator onto a multi-reference wave function. The latter, however, applies a separate cluster operator to each configuration of the multi-determinant wave function. ic-MRCC has a natural extension to wave functions encoded as MPSs, since the form of the cluster operator does not depend on the number of terms in the CI expansion of the wave function. Nevertheless, ic-MRCC has not yet been married with DMRG. Conversely, the requirement of having a separate cluster operator for each CI elements renders the extension of Jeziorski-Monkhorst CC theory to DMRG non-trivial. In fact, a wave function encoded as MPS can be virtually expanded in terms of an infinite number of basis functions.

A different strategy has instead been employed to couple CC with large-scale CI schemes, i.e. to express the wave function as in single-reference CC and to include multi-configurational effects by calculating the amplitudes involving strongly correlated orbitals from a CI (or MPS) wave function. For example, DMRG has been coupled with tailored CC with singles and doubles excitations (CCSD)\cite{Bartlett2006_TCC-Original} following this idea. In tailored CCSD, orbitals are partitioned in active, inactive and virtual as in multiconfigurational SCF (MC-SCF) approaches. The amplitudes associated to the single and double excitations within the active space are then extracted from a CAS (or DMRG\cite{Veis2016_TailoredCC-DMRG}) wave function. The remaining amplitudes are then optimized as in standard, single-reference CC by keeping the single- and double-excitation ones within the CAS fixed. The main advantage of tailored CC over its multi-reference counterpart is the computational cost, which is comparable to that of single-reference CCSD calculations. As already mentioned in the original paper,\cite{Bartlett2006_TCC-Original} even if a part of the amplitudes is obtained from multi-reference wave functions, tailored CC still represents a single-reference CC approach and this limits its accuracy for systems displaying strong static correlation, for which, however, the efficiency of DMRG is maximal. Moreover, the relevance of triple and higher-order excitations, that are neglected in tailored CCSD, has not been assessed yet.

Multi-reference methods can be combined with CC within the so-called externally corrected CC methods. These schemes extract high-order amplitudes for orbitals that are included in an active space from a CAS-SCF wave function, and optimize the full set of singles and doubles amplitudes, both for active and inactive orbitals, in the presence of these triple and quadruple excitations with standard CC.\cite{Paldus1997_ExternallyCorrected,Piecuch1999_ActiveSpace-CAS} A similar strategy has been employed by Piecuch and co-workers to combine CC including up to quadruple excitations (CCSDTQ) and FCIQMC. Unlike tailored CC, the high-order excitations are obtained from a multi-reference method, and not the low-order ones. For FCIQMC, it was shown that the amplitude of the triple and quadruple excitations obtained from a partially-converged FCIQMC calculation already provide nearly converged CCSDTQ energies.\cite{Piecuch2018_Stochastic-CC} Therefore, the same may hold true also for partially converged MPSs, obtained with a low bond dimension $m$. We note that FCIQMC has been combined by Piecuch and co-workers with CC also to automatize the so-called moment correction-based CC, that performs a CCSDTQ calculation by including triples and quadruples amplitudes only for excitations in the active space and estimates the effect of the remaining amplitudes by the so-called moment correction. As any multi-reference method, the accuracy of such a scheme depends strongly on the selection of the active space. To alleviate this problem, Piecuch and co-workers proposed to extract the predominant triples and quadruples amplitudes to be included in the exact CCSDTQ calculation from a partially converged FCIQMC propagation.\cite{Piecuch2017_Stochastic-CC,Piecuch2019_Stochastic-EOMCC} FCIQMC is, therefore, a driver that identifies the most relevant high-order cluster amplitudes to be included in the CC expansion, while the other ones are treated approximatively.

Canonical transformation (CT) theory\cite{Yanai2006_CanonicalTransformation} differs from CC. Whereas the wave function is parametrized using the same exponential operator as in unitary coupled cluster,\cite{Bartlett1989_Unitary-CC,Taube2006_Unitary-CC} the commutators entering the amplitude equations are then approximated by keeping only one- and two-particles operators in the Mukherjee and Kultzenigg generalized normal-ordered Hamiltonian.\cite{Mukerjee1997_GeneralizedNormalOrdering} Three- and higher-order reduced density matrices are approximated through a cumulant expansion.\cite{Yanai2007_CT-ExtedendNormalOrdering} The main advantage of CT over MRCC is the need to compute one- and two-body reduced density matrices only, still including higher-order reduced density matrices in an approximated way. However, higher-order reduced density matrices are approximated in terms of the one- and two-particle ones. This corresponds to retaining only the low-order contribution to the so-called cumulant approximation. Such approximation is accurate for single-reference systems, but it is known to converge much slower for strongly correlated systems,\cite{Kohn2012_DensityCumulant} that are however the cases for which the efficiency of DMRG is maximal.

\subsection{DMRG-DFT hybrid approaches.}

Short-range dynamical correlation may be considered by combining DMRG with DFT to alleviate the problem of the Coulomb cusp and introduce an approximated DFT-based correlation potential. A common hurdle of all methods combining DFT with wave function theories (WFTs)\cite{Gagliardi2016_pDFT-Review} is the so-called double counting problem. Any multi-configurational method will include, besides the static correlation energy, also part of the dynamical one. This second portion of the correlation energy should then not be included in the subsequent DFT calculation. There is, however, no exact definition of static and dynamical correlation energy and a quantification of this missing part of correlation is, therefore, not trivial. This double-counting problem can be avoided\cite{Savin1997_srDFT,Toulouse2004_srDFT-Theory} by partitioning the electron-electron interaction through range separation. In this way, the short-range part of the interaction can be included in the DFT treatment and the long-range part in the wave function-based calculation. The resulting theory, known as short-range DFT (sr-DFT) long-range wave function theory\cite{Fromager2007_srDFT-Original} is formally exact and does not suffer from the double-counting problem by construction. However, it requires the knowledge of a universal short-range exchange-correlation functional. Standard functionals, designed to capture all electron correlation, cannot be applied for this purpose and new functionals must be devised. The intrinsic approximation of this universal short-range exchange-correlation functional makes the accuracy of sr-DFT functional-dependent, which is a major limitation over the other approaches presented above. However, one needs to keep in mind that multi-reference perturbation theory is usually applied to second order and therefore not of ultimate accuracy. By contrast, an advantage of sr-DFT approaches is that they hardly require additional computational effort on top of the multi-configurational calculation. The accuracy of sr-DFT depends also the range-separation parameter that partitions the Coulomb interaction in a long-range part and a short-range one. Giner {\it et al.} recently suggested to choose this parameter based on the difference between the exact Coulomb kernel and the one that is obtained from a correlated wave function.\cite{Giner2018_srDFT-BasisSet} Since, in the limit of $r_{12} \rightarrow 0$, the former diverges while the latter is finite, Giner {\it et al.} proposed to choose the range-separation parameter so that the long-range part of the Coulomb interaction equals the Coulomb kernel obtained from an approximated wave function. The resulting range-separation parameter is, therefore, basis set-dependent and, as expected, tends to 0 in the complete basis set limit. Moreover, it is not a constant, but instead a function of the coordinates. Even if this scheme has been coupled, up to now, only with selected CI,\cite{Loos2019_srDFT-BasisSet} it can be applied to any correlated method for which the pair density is available, therefore including DMRG. We have combined sr-DFT with MPSs.\cite{Hedegard2015_DMRG-srDFT} Interestingly, treating part of the electron correlation with DFT accelerates the convergence rate of DMRG with respect to the bond dimension $m$ through a regularization of the active orbital space.\cite{Hedegard2015_DMRG-srDFT}

Fromager proposed to partition electron correlation in the orbital space.\cite{Fromager2015_CASSCF-DFT} A subset of the full set of orbitals is treated with wave function-based approaches (including DMRG\cite{Senjean2018_SiteOccupation-BetheAnsatz}) and their interaction with the remaining orbitals is described with DFT. Unlike standard DFT, where it is a function of the coordinates only, the density becomes orbital-dependent as well. As discussed in Ref.~\citenum{Fromager2015_CASSCF-DFT}, a consistent definition of exchange-correlation functionals in this framework requires the design of functionals of the orbital occupation, in place of the density. Exact orbital occupation-dependent functionals can be derived for simple models, such as Hubbard Hamiltonians.\cite{Fromager2015_CASSCF-DFT} More recently, a strategy to extend local density approximation functionals to this framework has been reported.\cite{Fromager2017_LDA-OrbitalOccupationDFT} The lack of well-established algorithms for the design of these new functionals has, however, limited the applications of this approach to model Hamiltonians only.

A third, formally different approach to combine WFT with DFT is pair-DFT (pDFT)\cite{Olsen2014_PairDFT-Original,Gagliardi2016_pDFT-Review} which does not rely on any partitioning of the electron-electron correlation, neither in real nor in orbital space. Instead, the energy expression contains the kinetic and Coulomb energies from a CAS-SCF reference calculation, whereas all exchange and correlation contributions are evaluated from functionals of the on-top density. This energy functional is then evaluated only once, from the one-body and on-top densities obtained for a CAS-SCF (or DMRG\cite{Gagliardi2018_pDFT-DMRG}) wave function. Evaluating the functional only once is a computational advantage of pDFT over sr-DFT, in which self-consistency between the WFT and DFT parts can be reached. However, also pDFT requires the design of new functionals, depending on both the one-body and the on-top density. Although in Ref.~\citenum{Olsen2014_PairDFT-Original} a strategy to design such functionals starting from standard ones was provided, we note that these new functionals should also include corrections associated to the kinetic energy, which is evaluated based on a truncated wave function and is, therefore, not exact. Moreover, self interaction and dispersion are notoriously difficult to be included in standard functional forms of density functionals. This makes the functional design highly non-trivial.

\section{DMRG for the nuclear Hamiltonian}
\label{sec:vDMRG}

\subsection{Vibrational DMRG}

Within the Born-Oppenheimer approximation, molecular vibrations are described in terms of the vibrational Schr\"{o}dinger equation,

\begin{equation}
  \mathcal{H}_\text{vib} | \Phi_\text{vib} \rangle 
    = \left[ \mathcal{T} \left( \bm{Q} \right) + \mathcal{V} \left( \bm{Q} \right) \right] 
  | \Phi_\text{vib} \rangle = E_\text{vib} | \Phi_\text{vib} \rangle \, ,
  \label{eq:VibScho}
\end{equation}
where $\mathcal{V} \left( \bm{Q} \right)$ is the potential energy surface (PES) operator obtained from the solution of the electronic Schr\"{o}dinger equation at different nuclear configurations $\bm{Q}$. Unlike the electronic Hamiltonian in Eq.~(\ref{eq:SQ_ElectronicStructure}), for which the interaction operators are known exactly (in the non-relativistic limit), the PES must be approximated, either with a Taylor series expansion about some reference geometry or with an $n$-mode expansion.\cite{Csaszar2012_Anharmonic-FF} Depending on the nature of this approximation, different second-quantized forms of the Hamiltonian in Eq.~(\ref{eq:VibScho}) are obtained, based either on the $n$-mode representation of the potential\cite{Christiansen2004_SecondQuantization,Wang2009_SQMCTDH} or on canonical quantization.\cite{Hirata2014_SecondQuantization} Both forms describe the vibrational motion in terms of Bose-Einstein statistics. We note that this would not be the case for the full molecular, \textit{i.e.} the pre-Born-Oppenheimer Hamiltonian, in which the symmetry would be different for bosonic and fermionic nuclei.\cite{Hammes2002_ProtonElectronCorrelation,Nakai2007_NuclearOribtalTheory,Matyus2012_MolecularStructurePreBO,Adamowicz2012_Review}

Most of the numerical methods designed to solve the electronic Schr\"{o}dinger equation have been extended to vibrational structure, including HF,\cite{Bowman1986,Gerber2002_CCVSCF,Hansen2010} CC,\cite{Christiansen2004_VCC} CI\cite{Carter1997_VCI,Bowman2003,Neff2009,Panek2014_LocalModes,Panek2016_AnharmonicBiomolecules} and perturbative\cite{Christiansen2003_VMP2,Barone2005,Bloino2012_GVPT2,Bloino2016_Review} approaches. The high computational cost of vibrational CI (VCI) has impeded its application to systems with more than 10 to 20 atoms so far and, as for the electronic-structure case, this problem can be alleviated by DMRG. A MPS/MPO-based formulation of DMRG for vibrational problems (referred to as vDMRG) was introduced by us.\cite{Baiardi2017_VDMRG} We note that, unlike the electronic Coulomb potential that is purely a two-body interaction, a many-body expansion of a PES contains three- and higher-order couplings. As we have already remarked and shown in the literature, the MPS/MPO-based formulation is the ideal framework for applying DMRG to such complex Hamiltonians. In parallel, a TT-based theory to calculate the eigenvalues of vibrational Hamiltonians was proposed.\cite{Oseledts2016_VDMRG} Two related strategies have been devised to reduce the computational cost of VCI: basis pruning techniques\cite{Avila2011_Pruning,Bowman2015_Pruning,Carrington2016_IterativePruning} and precontraction schemes.\cite{Bowman1991_Precontraction,Carrington2002_Precontraction} Methods of the first class reduce the computational effort of VCI by including only a subset of the full configurational space in the CI expansion. Conversely, precontraction schemes divide the vibrational degrees of freedom in different subsets. The vibrational Schr\"{o}dinger equation is first solved for each subset, neglecting the coupling between them. The basis for the final VCI calculation is then constructed from the eigenfunctions of these local Schr\"{o}dinger equations.\cite{Bowman1991_Precontraction,Carrington2002_Precontraction} DMRG combines the advantages of both schemes. As in pruning algorithms, the CI expansion involves a reduced basis, constructed iteratively to give the best approximation of the exact wave function in a least-squares sense. Moreover, the full diagonalization is replaced with the solution of a series of monodimensional eigenvalue problems. A similar strategy is followed by precontraction schemes, even if in DMRG no partition of the vibrational degrees of freedom is needed.

We conclude by noting that DMRG has also been applied to the solution of the rotational Schr\"{o}dinger equation.\cite{Roy2018_RotationalDMRG} In this respect, the inclusion of vibro-rotational contribution in vDMRG has not been explored yet and is required to match high-accuracy experimental data.

\subsection{Vibrational correlation in vDMRG}

The problem of recovering dynamical correlation is not limited to electronic structure problems: vDMRG suffers from the same limitation. The distinction between dynamical and static correlation has been, however, much less discussed in the literature in the context of vibrational structure. Static vibrational correlation energy can be defined, by analogy with its electronic counterpart, as the portion of the total vibrational correlation energy associated with the absence of a predominant configuration in the nuclear wave function expansion. Any molecule displaying a double-well potential along an inversion coordinate will feature strong static correlation. The exact vibrational wave function associated with the inversion coordinate is delocalized on both sides of the well, and therefore, it cannot be described in terms of one configuration localized on a single reference geometry. Conversely, if the vibrational wave function is accurately represented in terms of harmonic oscillator eigenfunctions, only dynamical correlation is present.

As for the electronic structure case, dynamical correlation is efficiently recovered by vibrational perturbation theory, based either on harmonic wave functions\cite{Barone2005,Bloino2012_AnharmonicProperties} or on a vibrational SCF reference.\cite{Christiansen2003_VMP2,Changala2016_VMP2CurvCoord} To increase the efficiency of vDMRG it is crucial to apply the variational correction only to the vibrational degrees of freedom displaying strong static correlation. The effect for the remaining modes may then be captured by perturbation theory. This procedure is the vibrational counterpart of the selection of a complete active orbital space. Low-frequency modes can be defined as strongly correlated, because they are, in most cases, strongly anharmonic, and therefore, they require a variational treatment. Any reaction-path Hamiltonian-based model relies on such a criterion.\cite{Page1988_RPH,Handy2002_H2O2,Tew2004}  A single, low-frequency mode is treated variationally, while all the other higher-frequency modes are treated either by harmonic approximation or by perturbation theory.

Energy-based criteria are usually not sufficient to detect strong correlation. Another indicator that has found extensive application in vibrational structure theories to detect strong correlation are so-called resonances. They are associated with nearly degenerate, strongly coupled states. The well-known Fermi resonances, in which the energy of an overtone is close to the one of a fundamental transition, lead to near-vanishing denominators in the perturbative energy expansion. In addition, other resonances, not associated to any divergence of the perturbative series, are known to be associated to strong static correlation. This is the case, for example, for the so-called Darling-Dennison resonances, involving near-degenerate fundamental (or overtone) transitions. Advanced algorithms to detect resonant terms have been proposed in the literature and applied to hybrid perturbative-variational schemes.\cite{Barone2005,Bloino2012_GVPT2,Krasnoshchekov2014_VVPT2,Piccardo2015_SymmetricTops}

In this respect, diagnostics obtained from one- and two-mode entropies, defined analogously to their electronic counterpart,\cite{Legeza2003_OrderingOptimization,Rissler2006_QuantumInformationOrbitals} could be used to identify strongly interacting modes and to quantify their interaction strength. For electronic structure calculations, the convergence of these descriptors with the bond dimension is much faster than that of the energy. For this reason, strongly interacting orbitals can be identified based on a fast, non-quantitative DMRG calculation that can reproduce a qualitatively correct wave function.\cite{Legeza2003_OrderingOptimization,Boguslawski2012_OrbitalEntanglement,Stein2016_AutomatedSelection} This also holds true for vDMRG. Accurate vDMRG calculations can then be carried out only for those strongly coupled modes, while the effect of the remaining ones may be recovered from perturbation theory. Perturbative corrections can be obtained as in the electronic-structure case, either from an explicit evaluation of sum-over-states expressions or by minimizing Hylleraas-type functionals.

\section{Time-dependent formulation of DMRG}
\label{sec:TD-DMRG}

DMRG is an optimization algorithm to minimize the energy functional for wave functions expressed in the MPS parametrization. This is equivalent to solve the time-independent (TI) Schr\"{o}dinger within the manifold of matrix product states. Solving the TI Schr\"{o}dinger equation is the most natural choice to target ground-state energies and low-order properties. Other quantities are, however, more easily obtained from the solution of the TD Schr\"{o}dinger equation. For instance, X-ray spectroscopy with TI methods requires the calculation of highly excited eigenstates of the electronic Hamiltonian and, as already highlighted in Sec.~\ref{sec:DMRG_Theory}, this task is much more complex compared to ground state calculations. Within a TD framework, X-ray spectra are obtained from the Fourier transformation of an appropriate time-dependent autocorrelation functions without the need of any diagonalization.\cite{DePrince2017_CoreCC-RealTime,Li2018_XRay-RealTime,Schuurman2018_XRay} Perturbation theories, including CASPT2 and NEVPT2, can also be reformulated in the time domain\cite{Sokolov2016_TD-NEVPT2} in terms of the Fourier transformation of time-dependent Green's functions that do not require the calculation of high-order reduced density matrices. Due to these advantages, which have been described in detail in a recent paper by Chan and co-workers,\cite{Mardirossian2018_ChoiceRepresentation} most of the electronic structure approaches have been reformulated in the time domain in the last years. In broad terms, the resulting algorithms are known as real-time electronic structure methods. Originally, the TD extension was developed for semiempirical\cite{Ghosh2017_TD-Semiempirical,Gagliardi2019_TD-Semiempirical} and DFT-based models.\cite{VanVoorhis2006_RealTimeTDDFT,Lopata2011_RealTimeTDDFT,Li2018_RealTime-Review} More recently, real-time extension of wave function-based approaches, including CAS-SCF\cite{Sato2013_TD-CAS,Sato2015_TD-RAS} and CC\cite{Nascimento2016_TDCC,DePrince2017_CoreCC-RealTime} have been proposed.

\subsection{Quantum dynamics with matrix product states}

The exact solution of the TD Schr\"{o}dinger equation is plagued, as its TI counterpart, from the curse of dimensionality. For this reason, real-time CAS-SCF simulations are currently feasible only for few-atom systems.\cite{Sato2013_TD-CAS} This limitation can be alleviated by MPSs introduced to the TD Schr\"{o}dinger equation (in Hartree atomic units)

\begin{equation}
  {\mathrm i} \frac{\partial | \Psi_\text{MPS}(t) \rangle}{\partial t} = \mathcal{H} | \Psi_\text{MPS}(t) \rangle \, .
  \label{eq:TDSE}
\end{equation}  

Eq.~(\ref{eq:TDSE}) is not an eigenvalue problem, and therefore cannot be solved with the ALS algorithm. We have already mentioned that applying any operator, such as the Hamiltonian $\mathcal{H}$, to an MPS ($\Psi_\text{MPS}$) increases its bond dimension. Hence, it follows from Eq.~(\ref{eq:TDSE}) that the bond dimension increases during the propagation. This leads to an increase of the computational time needed to evaluate $\mathcal{H} \Psi_\text{MPS}(t)$ as the propagation evolves and to a high computational cost, especially for long propagations.\cite{Luo2002_Comment-TDDMRG} To limit the computational demands, the bond dimension of $\mathcal{H} \Psi_\text{MPS}(t)$ can be kept fixed by building the renormalized basis at the beginning of the simulation and keeping it fixed during the propagation.\cite{Cazalilla2002_TD-DMRG_FixedBasis} However, the renormalized basis is optimized to represent the initial wave function, but its accuracy deteriorates with increasing time. To solve this problem, the basis function can be updated at each time step,\cite{Feiguin2005_Adaptive-TDDMRG,Sokolov2017_TDDMRG-Perturbation,Ronca2017_TDDMRG-Targeting,Ren2018_TDDMRG-Temperature} to keep the accuracy fixed during the whole propagation. These algorithms are usually known as adaptive TD-DMRG.

A second major challenge associated with TD-DMRG is related to the numerical integration of the differential equation itself. Formally, its solution reads

\begin{equation}
  \Psi_\text{MPS}(t) = e^{- {\mathrm i} \mathcal{H}t} \Psi_\text{MPS}(0) \, ,
  \label{eq:Propagator}
\end{equation}
where $e^{-i\mathcal{H}t}$ is the propagator (i.e, the time-evolution operator). To evaluate efficiently Eq.~(\ref{eq:Propagator}), the exponential operator must be encoded as an MPO, as in Eq.~(\ref{eq:MPO}). For Hamiltonians containing nearest-neighbor interactions only,\cite{Suzuki1976_TrotterApproximation} such representation is obtained by approximating the propagator through a Suzuki-Trotter splitting. The resulting theory, known as time-evolving block decimation (TEBD),\cite{Vidal2004_TEBD} has been successfully applied to nearest-neighbour Hamiltonians, such as the Hubbard one, but is not general enough for QC Hamiltonians that show long-range interactions.

Alternatively, the TD Schr\"{o}dinger equation can be solved with numerical methods such as the Runge-Kutta\cite{Feiguin2005_Adaptive-TDDMRG,Ronca2017_TDDMRG-Targeting,Sokolov2017_TDDMRG-Perturbation,Ren2018_TDDMRG-Temperature} or the Lanczos schemes,\cite{Frahm2019_TD-DMRG_Ultrafast} adapted to MPSs. The fourth-order Runge-Kutta scheme is mostly applied in conjunction with the adaptive TD-DMRG introduced above because it calculates the wave function after a time step $\Delta t$ from the wave function at the initial time $t$ and at times $t + \Delta t/ 4$, $t + \Delta t/2$ and $t + 3 \Delta / 4$. These intermediate wave functions can be employed to determine the optimal renormalized basis for the final wave function at time $t + \Delta t$.\cite{Feguin2006_Adaptive-TDDMRG,Sokolov2017_TDDMRG-Perturbation,Ronca2017_TDDMRG-Targeting,Ren2018_TDDMRG-Temperature}

\begin{figure}
  \centering
  \includegraphics[width=.35\textwidth]{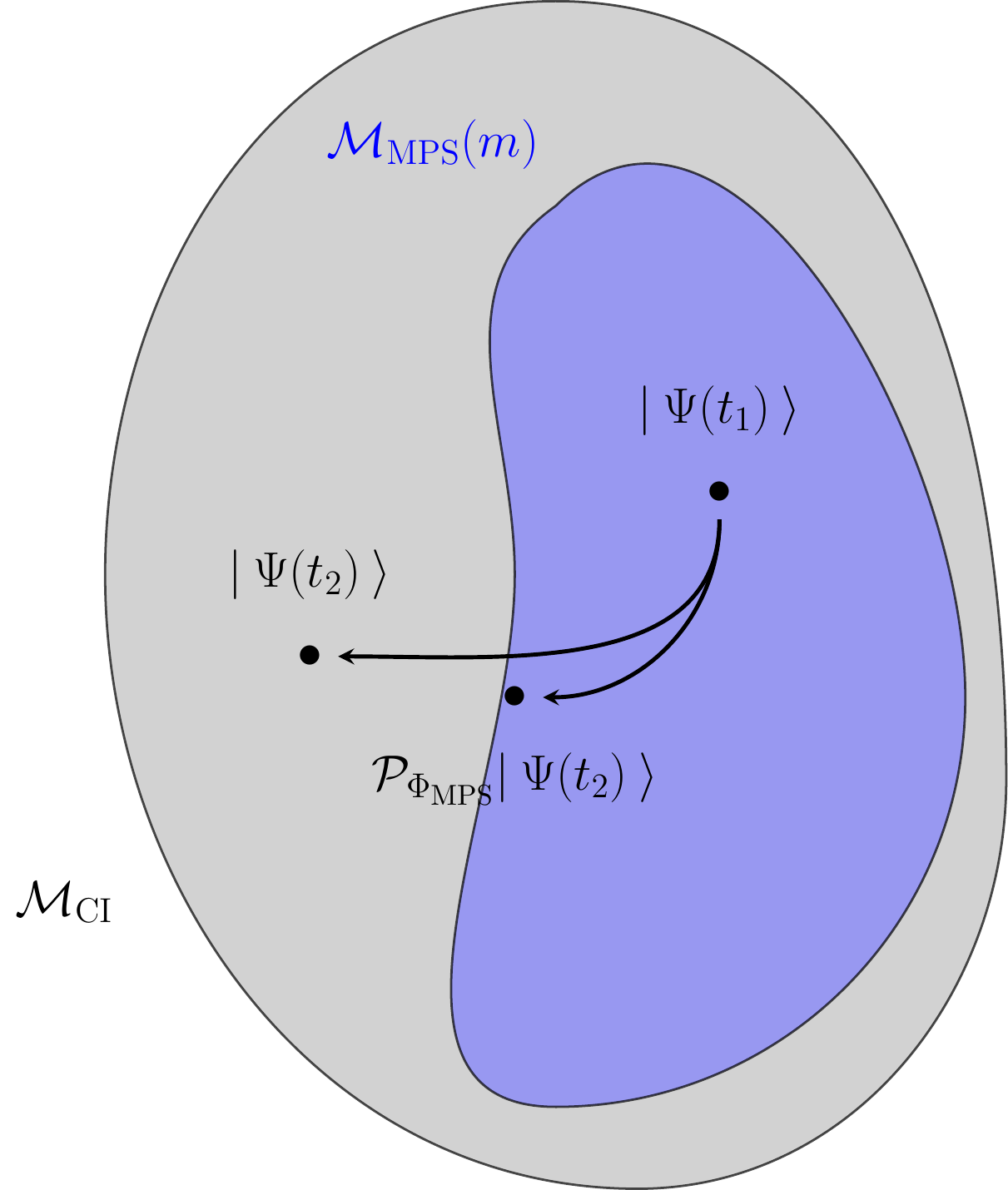}
  \label{fig:Propagation}
  \caption{Graphical representation of the tangent-space TD-DMRG approach. The gray set represents the full-CI space ($\mathcal{M}_\text{CI}$) and the blue space represents the tangent space to the manifold of the MPS with a fixed bond dimension $m$ ($\mathcal{M}_\text{MPS}(m)$) calculated at the MPS at time $t_1$ ($| \Phi (t_1) \rangle$). The exact wave function at a successive time $t_2 > t_1$ ($| \Phi (t_2) \rangle$) is not an element of the tangent space. Its projected counterpart ($\mathcal{P}_{\Phi_\text{MPS}}| \Phi (t_2) \rangle$) is the best approximation of $| \Phi (t_2) \rangle$ in $\mathcal{M}_\text{MPS}(m)$.}
\end{figure}

The TD Schr\"{o}dinger equation can be also recast as a variational problem by applying the well-known TD Dirac-Frenkel variational principle (TDVP).\cite{Moccia1973_TDVP} Within this framework, the time evolution of an MPS is determined by minimizing the following functional:

\begin{equation}
  F\left[ \Psi_\text{MPS}(t), t \right] 
    = \left\| {\mathrm i} \frac{\text{d} \Psi_\text{MPS}(t)}{\text{d} t} 
    - \mathcal{H} \Psi_\text{MPS}(t) \right\|^2 \, ,
  \label{eq:TDVP_Minimization} 
\end{equation}
where the minimization is performed over the MPSs with a fixed bond dimension $m$. The resulting propagation will be approximate, since an exact solution of Eq.~(\ref{eq:TDSE}) would lead to a continuous increase of $m$. However, as in standard DMRG, the full-CI limit is recovered by systematically increasing $m$. Eq.~(\ref{eq:TDVP_Minimization}) can be recast as

\begin{equation}
  {\mathrm i} \frac{\partial | \Psi(t)_\text{MPS} \rangle}{\partial t}
    = \mathcal{P}_{\Psi_\text{MPS}} \mathcal{H} \Psi(t)_\text{MPS} \, ,
  \label{eq:TDVP}
\end{equation}
where $\mathcal{P}_{\Psi_\text{MPS}}$ is the projector onto the manifold of all possible MPSs with bond dimensions $m$. As shown in Figure~\ref{fig:Propagation}, the projector ensures that the wave function is described as an MPS of bond dimension $m$ during the whole propagation. This projector can be expressed as a sum of site terms, as has been recently demonstrated in the context of TT theory.\cite{Holtz2012_ManifoldTT,Lubich2014_TimeIntegrationTT} The propagator of Eq.~(\ref{eq:Propagator}) can, therefore, be factorized as a product of site terms, and its action on an MPS can be calculated by applying the terms sequentially, as in ALS minimization.\cite{Zaletel2015,Haegeman2016_MPO-TDDMRG} A similar tangent space-based scheme has been recently introduced by Bonfanti and Burghardt for the multi-configurational time-dependent Hartree (MCTDH) scheme.\cite{Bonfanti2018_MCTDH-ProjectorSplitting} So far, this tangent-space formulation of TD-DMRG has been applied to model vibrational Hamiltonians.\cite{Borrelli2017_TDDMRG,Batista2017_TT-SOFT,Tempelaar2018_Holstein-TDDMRG,Shuai2019_TDDMRG-GPU} However, due to its generality, the framework can be applied to \textit{ab initio} Hamiltonians as well. The application of TD-DMRG to time propagations of a nuclear wavepacket on vibronic Hamiltonians have been introduced by us\cite{Baiardi2019_TDDMRG} enables one to simulate photochemical processes with DMRG for systems with more than 20 vibrational degrees of freedom. In this respect, TD-DMRG could constitute an efficient alternative to the MCTDH algorithm,\cite{Meyer2011_MCTDH-Review} which is currently the reference method for quantum dynamics simulations. MCTDH can be interpreted as the time-dependent vibrational analog of CAS-SCF since during the propagation both the CI wave function and the modals are optimized simultaneously. As TI CAS-SCF, MCTDH suffers from the curse of dimensionality and its computational cost scales exponentially with the number of degrees of freedom. A multilayer formulation of MCTDH (ML-MCTDH) has been introduced\cite{Manthe2008_MLMCTDH-Original,Wang2009_SQMCTDH,Meyer2013_MLMCTDHVibronic,Burghardt2013_ChargeSeparation-MLMCTDH,Burghardt2013_ML-G-MCTDH} to limit this increase, where vibrational coordinates are coupled according to a hierarchical contraction scheme. In Section~\ref{sec:DMRG_Theory} we discussed how some multi dimensional generalizations of the MPS parametrization can be interpreted as a hierarchical extension of the MPS. In this respect, we believe that ML-MCTDH is not an alternative to TD-DMRG, but both approaches can rather be combined to improve their respective efficiency.

A fundamental limitation of any TD-DMRG approach is that there is no guarantee that a wave function can be represented as a compact MPS during a propagation. This makes the assessment of the convergence of TD-DMRG not trivial for long-time simulations, as has been discussed by Reichmann and co-worker for spin Hamiltonians.\cite{Kloss2018} We also noted that,\cite{Baiardi2019_TDDMRG} if the bond dimension $m$ is adapted dynamically during the propagation to get a constant truncation error,\cite{Legeza2003_DynamicalBlockState} the bond dimension increases linearly with time. This means that fixing $m$ introduces and error that grows linearly with time. However, we observed\cite{Baiardi2019_TDDMRG} that some observable, such as autocorrelation functions, converge quickly with $m$. Absorption spectra are mostly governed by short-time propagations and, therefore, the impact of the long-time error is expected to be small.

Legeza and co-workers have proposed to optimize the local basis during the propagation to improve the accuracy of TD-DMRG. Following an algorithm originally introduced to improve the convergence of TI-DMRG,\cite{Legeza2016_OrbitalOptimization,Krumnow2019_FermionicOrbitalOptimization} the basis is optimized by minimizing the entanglement entropy of the MPS,\cite{Legeza2019_OrbitalOptimization-TD} that grows with the bond dimension $m$. The optimization is carried out by applying a unitary transformation to the local basis of two neighbouring sites after each micro-iteration of a DMRG sweep.

Also in MCTDH\cite{Worth2008_MCTDH-Review,Meyer2011_MCTDH-Review}  the local basis is optimized during the propagation, but the optimization is realized in a substantially different way than in Ref.~\citenum{Legeza2019_OrbitalOptimization-TD}. The local basis is expressed as linear combination of a larger basis set, referred in the following as primitive basis, in the same way as CAS-SCF molecular orbitals are expressed in terms of an atomic basis. The coefficients of this linear combination are optimized during the propagation applying the TDVP. Therefore, the vector space spanned by the local basis, that is a subset of the space spanned by the primitive basis, changes dynamically during the propagation. This is not true for the algorithm described above,\cite{Legeza2019_OrbitalOptimization-TD} in which transformations are applied \textit{within} the local basis to obtain a compact MPS. Kurashige\cite{Kurashige2018_MPS-MCTDH} proposed a scheme to couple this MCTDH local basis optimization with TD-DMRG based on the theory of Ref.~\citenum{Bonfanti2018_MCTDH-ProjectorSplitting}. Even if applications of this scheme are still limited to very small systems, but results suggest that also this scheme could improve significantly the efficiency of TD-DMRG. We conclude by noting that the two optimization schemes introduced above can be, in principle, coupled to select the best DMRG local basis with the MCTDH-based optimization\cite{Worth2008_MCTDH-Review,Meyer2011_MCTDH-Review} and applying the entanglement minimization\cite{Legeza2019_OrbitalOptimization-TD} to obtain the most compact MPS parametrization within this basis set.

Interestingly, even if the TI formulation of DMRG has been applied mostly to electronic structure problems, applications of its TD counterpart have often been limited to vibrational Hamiltonians.\cite{Ren2018_TDDMRG-Temperature,Baiardi2019_TDDMRG} TD-DMRG for \textit{ab initio} electronic structure Hamiltonians enables one to simulate electron dynamics. Even if the extension of MCTDH to electronic processes has been known for more than ten years,\cite{Saalfrank2005_MCTDH-Electrons} its application has been limited by the absence of experimental reference data. Owing to the impressive development of attosecond spectroscopic techniques,\cite{Corkum2007_AttosecondScience,Worner2015_Iodoacetylene,Lara-Astiaso2018_AttosecondReview} it is now possible to probe electron dynamics in real time. There is therefore a need for accurate electronic structure methods supporting the interpretation of attosecond spectra. Currently, the only multi-reference theories applicable to time-depdendent processes are TD-CAS-SCF\cite{Sato2013_TD-CAS,Sato2015_TD-RAS} and TD-CI.\cite{Schlegel2014_StrongFieldIonization,Peng2018_TDCI,Li2018_RealTime-GUGA_CI,Evangelista2019_Adaptive-TDCI} Their high computational cost has limited them so far to few-atom molecules. The extension of the TD-DMRG framework designed by us for vibrational and vibronic problems\cite{Baiardi2019_TDDMRG} to the electronic Hamiltonian would extend the range of applicability of TD-CAS-SCF, allowing to study transition metal complexes or complex biomolecules.

\subsection{Imaginary-time propagation of matrix product states}

An interesting further development of TD-DMRG is its extension to imaginary-time propagation. It is well-known that expressing the time in the complex domain allows the study thermal ensembles\cite{Ren2018_TDDMRG-Temperature} and open quantum systems.\cite{Aspuru2010_TDDFT-OpenSystems} The inclusion of temperature effects is particularly relevant for vibrational Hamiltonians, because the energy of nuclear motions is comparable to the thermal energy at room temperature. Nevertheless, in presence of low-lying electronic states, temperature effects might become relevant also for electrons. For this reason, some electronic structure theories have been generalized to include temperature effects.\cite{Chan2018_FiniteTemperature-CC} The same strategy can be followed to generalize TD-DMRG to thermal ensembles.\cite{Ren2018_TDDMRG-Temperature}

The particular case in which the time variable is a purely imaginary number corresponds to the limit of zero temperature. In that case, the TD propagation becomes equivalent to a ground-state optimization. This idea is exploited in diffusion Monte Carlo,\cite{Tolouse2016_DMC-Review} as well as in FCIQMC,\cite{Alavi2009_FCIQMCOriginal} where the propagation is replaced by a stochastic dynamics. Imaginary-time propagation of MPSs is, therefore, an alternative to the standard, ALS-based optimization, as we proved for vibrational Hamiltonians.\cite{Baiardi2019_TDDMRG} This alternative is particularly appealing for general tensor network states, for which ALS is not available. As already recalled above, the optimization is the bottleneck of any calculation involving tensor network states. Therefore, imaginary time propagation could lead to a speed-up of tensor network states optimization compared with the currently available algorithm. The extension of the Dirac-Frenkel principle to general tensor networks would, however, require a closed-form expression for the projection operator onto the tangent space, which is not known.

A significant difference between ALS-based optimization techniques and the imaginary-time propagation is that the former are variational, while the latter are projective. As pointed out by Alavi and co-workers,\cite{Alavi2018_FCIQMC-Transcorrelated} projective optimization techniques are particularly appealing when coupled to \quotes{dressed} non-Hermitian Hamiltonians, obtained by non-unitary transformation of Eq.~(\ref{eq:SQ_ElectronicStructure}) to partially include correlation effects in the definition of the Hamiltonian itself. Among them, the trans-correlated Hamiltonian introduced by Boys and Handy\cite{Handy1968_TranscorrelatedHamiltonian} automatically includes in its definition a Jastrow-like factor without the need of considering it explicitly in the wave function. The applications of the transcorrelated Hamiltonian to quantum chemical problems has been hampered by the fact that it is not Hermitian, and therefore, its eigenfunctions are not well-defined.\cite{Tenno2000_Transcorrelated,Luo2010_VariationalTranscorrelated,Yanai2012} This issue can be circumvented with projection-based optimization techniques, such as FCIQMC,\cite{Alavi2009_FCIQMCOriginal} which do not require any modification when dealing with non-Hermitian operators. Imaginary-time propagation would, therefore, pave the route towards the coupling of DMRG with dressed Hamiltonians, including the trans-correlated one.

\section{Application of DMRG to quantum chemical problems}
\label{sec:applications}

\subsection{DMRG studies of complex multireference systems}

The first implementations of DMRG to the electronic structure Hamiltonian were tested on HHeH,\cite{Ciofini2000_DMRG} LiF,\cite{Legeza2003_DMRG-LiF} H$_2$O,\cite{Chan2003}  N$_2$\cite{Chan2004_NitrogenDMRG} and CsH,\cite{Moritz2005_DKH-DMRG} all molecules with less than 5 atoms, for which full CI calculations are still feasible. The availability of CI results made these systems ideal to study the convergence of DMRG. Subsequent applications were mostly limited to quasi-onedimensional molecules, for which the efficiency of DMRG should be best (ignoring the long-range Coulomb interaction) and fully converged results are obtained with values of $m$ of the order of magnitude of 100. Typical examples include linear hydrogen chains\cite{Hachmann2006_QuadraticSCF,Ronca2017_TDDMRG-Targeting} as well as $\pi$-conjugated organic systems,\cite{Yanai2010_Polycarbene} and in particular polyenes.\cite{Ghosh2008_OrbitalOptimization,Dorando2009_AnalyticalResponseFunction,Hu2015_DMRGExcitedStateOptimization,Wouters2016_DMRG-CASPT2} For these systems, DMRG can converge ground-state energies for active spaces with up to 100 orbitals, a size not reachable by standard CAS algorithms.

For more complex molecules, a higher bond dimension $m$ is needed to converge DMRG. However, for many (if not most) applications a value of $m$ lying between 1000 and 10000 will yield sufficiently accurate converged energies.
We argued in 2008\cite{Marti2008_DMRGMetalComplexes} that relative energies of compact molecules such as transition metal complexes can be obtained with DMRG, which initiated the application of DMRG as a reference method in this field. A prominent example is the DMRG study of the electronic properties of synthetic Fe-S clusters, which are found as active sites in metalloenzymes.\cite{Sharma2014_IronSulfur-DMRG} For these systems, the full-variational energy of the lowest 10 electronic states could be obtained from DMRG calculations with $m$=4000 and including up to 30 orbitals. Also the Mn$_4$CaO$_5$ cluster, which is buried in photosystem II and responsible for oxygen production on Earth, was a target for DMRG calculations.\cite{Kurashige2013_MnCaCluster-DMRG}

However, these works suffer from common limitations. First, the convergence is assessed for a given CAS, but the choice of the orbitals to be included in the CAS is not discussed. Furthermore, dynamical correlation is not included at any level. This limit is mentioned, for example, in Ref.~\citenum{Kurashige2013_MnCaCluster-DMRG}, where it was shown that DMRG calculations provide correct energy ordering for the first 10 excited states of the Mn$_4$CaO$_5$ cluster, but do not correctly reproduce the absolute energies. By virtue of recent developments of different perturbation theories combined with DMRG, the most recent applications employ DMRG-CASPT2\cite{Chalupski2014_DMRGPT2-Catalysis,Pierloot2017_NiFeCatalisys-DMRG,Pierloot2018_BimetallicIron-DMRG} or DMRG-NEVPT2\cite{Freitag2017_DMRG-NEVPT2} to reliably include dynamical correlation effects. Among most advanced applications, we mention here the work of Yanai and co-workers\cite{Chalupski2014_DMRGPT2-Catalysis} where DMRG-CASPT2 is applied to the study of the catalytic dehydrogenation of alkenes by the desaturase enzyme. This study shows that PT can account for corrections of up to 100 kJ/mol to the pure DMRG result and its inclusion is, therefore, crucial for a correct identification of the reactive intermediates of the catalytic reaction. Similar considerations are reported also a recent DMRG study of spin-crossover metal complexes,\cite{Freitag2017_DMRG-NEVPT2,Pierloot2018_BimetallicIron-DMRG} which are usually characterized by small singlet-triplet energy gaps. The inclusion of dynamical correlation effects is again crucial to correctly reproduce this energy gap. The reliability of modern DMRG-PT2 approaches makes them one of the \quotes{gold-standard} methods to obtain reference data for multi-configurational systems to design and to test new DFT functionals for the study of transition metal complexes.\cite{Chalupski2014_DMRGPT2-Catalysis,Pierloot2017_NiFeCatalisys-DMRG}

\subsection{Automatic selection of active orbital spaces}

As is true for any multi-configurational method defined for a chosen orbital space, the accuracy of DMRG strongly depends on the definition of the CAS. It is a natural desire deeply rooted in scientific objectivism to make this choice based on rigorous criteria without any human interference. The actual practice, however, is quite different and expert knowledge is considered to be key. However, the fact that DMRG can address very large orbital spaces with iteratively increasing accuracy holds a key to this problem. In turn, DMRG calculations can be fully automated, therefore making it as black-box as single-reference methods, such as DFT and CC. 

Hence, a descriptor measuring the degree of entanglement of states defined on a subset of orbitals must be defined, based on which the most strongly entangled orbitals can be chosen for a CAS in a fully automatic way. For an orbital selection scheme to be of universal applicability, some requirements must be met: 
(1) An orbital selection scheme should be agnostic with respect to the type of orbital basis from which orbitals shall be selected, 
(2) reliable objective (absolute) criteria for the orbital classification are necessary,
(3) as few general parameters (thresholds) as possible should be set to demonstrate general applicability (e.g., a single decision criterion is desirable),
(4) the scheme must be able to inspect all relevant orbitals for the process to be described (e.g., all valence orbitals if reaction energies are a target so that the selection scales with the molecular size),
(5) it must work for any type of molecule (i.e., it should be agnostic with respect to elements from periodic table),
(6) it must work along reaction coordinates and for excited states to be useful in applications on chemical processes,
(7) absolutely no manual interference in the whole selection process must be required if the scheme shall be called automated,
(8) the selection scheme should enable fully automated CAS-SCF or DMRG-SCF calculations that require as little input as is required for DFT or CC calculations, and
(9) it can be beneficial to have an absolute diagnostic for measuring static electron correlation available.

As already discussed in the context of orbital ordering optimization, measures obtained from quantum information theory are particularly well-suited for quantifying orbital correlations. In particular, a reliable metric is the single-orbital von Neumann entropy $s_i(1)$, which measures the deviation of a spatial-orbital sub-state from one of the four pure states of a spatial orbital\cite{Legeza2003_OrderingOptimization,Legeza2004_DataCompression-DMRG,Rissler2006_QuantumInformationOrbitals}

\begin{equation}
s_i(1) = -\sum_{\alpha=1}^4 w_{\alpha,i} \ln \left( w_{\alpha,i} \right) \, ,
\label{eq:SingleOrbitalEntropy}
\end{equation}
where $w_{\alpha,i}$ are the eigenvalues of the one-orbital reduced density matrix. The two-orbital entropy $s_{ij}(2)$ can be defined analogously to Eq.~(\ref{eq:SingleOrbitalEntropy}). The mutual information $I_{ij}$ between orbitals $i$ and $j$ is defined in terms of $s_i(1)$ and $s_{ij}(2)$ as\cite{Legeza2003_OrderingOptimization,Rissler2006_QuantumInformationOrbitals,Boguslawski2012_OrbitalEntanglement}

\begin{equation}
I_{ij} = \frac{1}{2} \left[ s_i(1) + s_j(1) - s_{ij}(2) \right] \left( 1 - \delta_{ij} \right) \, .
\label{eq:MutualInformation_Definition}
\end{equation}

Eq.~(\ref{eq:MutualInformation_Definition}) has the following intuitive interpretation: if orbitals $i$ and $j$ are independent, \textit{i.e.}, not entangled, the two-body entropy $s_{ij}(2)$ is just the sum of the one-orbital entropies, $s_i(1) + s_j(1)$, hence $I_{ij} = 0$. Conversely, in presence of orbital interaction, the entanglement of the pair $(i,j)$ decreases compared to the rest of the system ($s_{ij}(2)$), and thus $I_{ij} > 0$. 

Legeza and co-workers \cite{Legeza2003_OrderingOptimization,Legeza2004_DataCompression-DMRG} introduced these entanglement measures calculated from a fast unconverged DMRG calculation in a {\it given} orbital space to prepare the ordering of orbitals on the one-dimensional DMRG lattice for a subsequent fully converged calculation.
It was later shown \cite{Boguslawski2012_OrbitalEntanglement,Boguslawski2013_OrbitalEntanglement-BondFormation} that orbitals responsible for large static correlation effects usually have a large single-orbital entropy and large mutual information. As convergence of qualitatively correct one- and two-orbital entropies is faster than convergence of the energy, a partially converged DMRG calculation, which is comparatively fast to carry out, can deliver reliable approximate entropy values (a potential failure of this procedure can be probed and corrected after the fully converged results in a selected CAS are obtained).\cite{Stein2016_AutomatedSelection} 
Based on this idea, we proposed an automated protocol for the {\it initial} selection of orbitals for the DMRG lattice (not just for its sorting), which can be fully automated. This automated protocol for the CAS definition is referred to in the following as \texttt{AutoCAS}.\cite{Stein2016_AutomatedSelection,Stein2016_DelicateBalance,Stein2017_OrbitalEntanglment,Stein2017_AutoCAS-Chemia,Stein2019_AutoCAS-Implementation} 
We implemented the \texttt{AutoCAS} selection algorithm in a graphical user interface available that is available free of charge from our webpages. It fulfills the requirements for truly automated orbital selection listed above.

First, the entanglement metrics are extracted from a partially converged DMRG calculation based on a large valence active space (if it is too large for a single-shot DMRG calculation, it can be efficiently disected with results patched together afterwards \cite{Stein2019_AutoCAS-Implementation}). Strongly entangled orbitals are then identified and included in a smaller CAS representing strong static electron correlation well. This CAS is then employed in fully converged DMRG calculations. The single-orbital entropy, normalized with respect to its maximum value among all orbitals of the active space, is sufficient to identify strongly entangled orbitals and the mutual information does not provide any additional insight. If applied to molecules already studied with post-HF methods, \texttt{AutoCAS} can lead to a different and more accurate definition of the CAS with respect to the works already available in the literature (see, for instance, the case of several metallocenes\cite{Stein2016_DelicateBalance}). The reliability of the one-orbital entropies only (rather than including also the mutual informaton, which would be easily possible) is due to the fact that it comprises information from the one-body, but also from the two-body reduced density matrices. Although the grand-canonical one-orbital reduced density matrix is easy to obtain in a DMRG program, it can also be obtained in traditional CAS-type calculation from the standard one- and two-body density matrices \cite{Tecmer2015_OrbitalEntanglement}. \texttt{AutoCAS} was shown to work well for systems, for which hardly any complete set of standard rules for orbital selection can be applied, such as a dinuclear Iridium catalyst.\cite{Stein2017_AutoCAS-IrCatalysis} However, its applicability range can be easily enlarged to include, e.g., i) excited states not governed by valence orbitals through the consideration of Rydberg-type orbitals in the orbital selection step, ii) core excitations in X-ray spectroscopy, and iii) magnetic orbitals in antiferromagnetic couplings. In fact, the last option was discussed in detail and shown to be a viable target for \texttt{AutoCAS} in a recent paper by Stein {\it et al.}\cite{Stein2019_AntiferromagneticCouplings}.

An alternative route to define automatically active spaces has been proposed based on natural-orbital occupation numbers (NOONs), which has been considered for a long a time (see, e.g., Refs. \citenum{Pierloot2010_Corroles,Pierloot2011_NOON,Keller2015_AutomatedSelection} and references cited therein). The difference between the NOONs of strongly- and weakly-correlated orbitals is, however, often much less pronounced than for single-orbital entropies. We have argued\cite{Stein2016_AutomatedSelection} that NOONs are less sensible to correlation because of their absolute values which cluster in two distinct regions, whereas orbital entropies show a broad spread and can be normalized with respect to the highest value found in a molecule under consideration. 

A major limitation of \texttt{AutoCAS} is that a DMRG calculation in the full valence space, even partially converged, may become prohibitive for more than 60--100 orbitals. This is a common limitation of any top-down selection scheme, including NOON-based ones, that rely on a full-valence correlated calculations. We recently presented an algorithm\cite{Stein2019_AutoCAS-Implementation} that calculates entanglement measures for active spaces with more than 100 orbitals by partitioning the active space into several subsets of orbitals and by carrying out DMRG calculations in these subspaces. 

In recent years, a large number of active space selection schemes have been proposed. However, none of them is truly generally applicable and fully automated according to the list of requirements given above. In the following, we discuss two schemes, which are related to our DMRG context here. One algorithm to choose an active space is the atomic valence active space (AVAS)\cite{Chan2017_AutomaticSelection} scheme introduced by Chan and co-workers. AVAS includes in the active space the HF orbitals with the largest overlap with a set of atomic orbitals that are known \textit{a priori} to give rise to strong correlation effects (such as the $d$ orbitals in a metal complex). AVAS is, however, not fully automatic since it is heavily dependent on the choice of the target atomic orbitals. The lack of generality is also the drawback of the scheme proposed by Sayfutyarova and Hammes-Schiffer,\cite{Sayfutyarova2019_AutoCAS-PiSystems} that is tailored to $\pi$-conjugated systems. 

Alternatively, Khedkar and Roemelt\cite{Khedkar2019_NEVPT2-AutoCAS} designed a bottom-up selection scheme that builds up an active space by identifying strongly correlated orbitals based on the natural occupation number obtained from strongly-contracted NEVPT2. The computational cost of this scheme is, however, governed by the expensive NEVPT2 step, which also limits its applicability to systems with up to about 30 orbitals.

\subsection{Embedding schemes}

The successful applications of the most recent DMRG formulations, possibly coupled with PT theories, to strongly correlation systems, paves the route towards even more advanced applications of DMRG. A necessary step to extend the range of applicability of DMRG is the coupling with embedding schemes, to target even larger systems, possibly in a complex environment. Most of the embedding schemes applied to CAS-SCF can be trivially extended to DMRG. This could be the case, for example, of the polarizable continuum model (PCM),\cite{Tomasi2005_Review} whose coupling with CAS-SCF calculations is available in the literature for more than 20 years.\cite{Barone1999_PCM-CASSCF} PCM describes efficiently non-polar solvents, but complex environments require more refined embedding techniques. The latter include, for example, wave function-in-DFT approaches (WFT-in-DFT), in which the relevant portion of the molecules is described with WFT, while the rest of the molecule is treated at the DFT level. The DFT density introduces an external potential to be included in the WFT-based treatment, while the WFT density modifies the energy functional of the DFT part, so that the two densities should be calculated self-consistently. Based on this idea, DFT-based embedding schemes have been coupled with several WFTs, including CC,\cite{Jacob2009_CCinDFT-Embedding} MP2\cite{Carter1998_MP2Embedding} and CAS-SCF.\cite{Wesolowski2008_WFTinDFT-Embedding,Carter2012_WFTinDFT-Embedding} As for PCM, also in this case the latter theory can be straightforwardly extended to DMRG.\cite{Reiher2015_DMRGinDFT-Embedding} One of the main challenge of WFT-in-DFT embedding schemes is the need of designing the so-called non-additive contribution to the kinetic energy functional. Embedding schemes introduced by Miller and Manby\cite{Manby2012_Embedding,Miller2015_EMFT,Muhlbach2018_Embedding} bypasses this problem, e.g., by adding a projection operator to the Kohn-Sham operator to enforce orthogonality between orbitals belonging to different subsystems. Originally developed for DFT-in-DFT, these embedding schemes have been extended to various WFTs, including MP2\cite{Manby2014_ProjectionEmbedding-MP2} and CC.\cite{Manby2012_OpenShell-DFTEmbedding,Manby2016_CCinDFT-CitrateSynthase}

A more detailed description of environmental effects can also be obtained with mixed quantum/classical mechanical (QM/MM) models, in which the environment is represented through classical point charges.\cite{Warshel1976_QMMM,Truhlar2007_QM-MM-Review,Thiel2009_QM-MM-Review} This atomistic description of the environment is a significant improvement over PCM, in which solvent effects are averaged. QM/MM methods can be broadly divided in non-polarizable and polarizable ones, the latter being more accurate owing to the inclusion of mutual polarization between the QM and the MM part. Among the various polarizable QM/MM approaches available in the literature,\cite{Kongsted2010_PolarizableEmbedding-ExcitedStates,Lipparini2011,Loco2016_PolarizableEmbedding} only the induced dipole theory introduced in Ref.~\citenum{Kongsted2013_PolarizableEmbedding-Review} has been extended to DMRG so far.\cite{Hedegard2016_PE-DMRG}

More recently, alternative schemes have been proposed, where the molecular system has been partitioned in orbital space instead of in coordinate space (see Ref. \citenum{Muhlbach2018_Embedding} and reference therein). The starting point of all these theories is a low-level mean-field calculation, such as HF, from which a set of orbitals is defined. The orbitals are then partitioned into different groups, each of which is treated at a different level of theory. The accuracy of these embedding schemes depends heavily on the partition of the system and on the \textit{a-posteriori} inclusion of coupling effects between different blocks of orbitals. At the lowest level the couplings can be simply neglected,\cite{Mata2008_EmbeddingLocalCorrelation} hence leading to separate, non-interacting electronic structure calculations. Improvements are obtained by including the effects of the low-level calculations on the higher-accuracy ones at the mean-field level, as recently accomplished for the MRCC in CAS-SCF embedding scheme.\cite{Kohn2018_Embedded-MRCC}

A conceptually different embedding strategy is DMET proposed by Knizia and Chan\cite{Chan2012_DMET,Chan2013_DMET-StrongCoupling} which relied on a mean-field embedding in its first version. Higher-accuracy calculations can then be performed on a smaller portion of the molecule (known as the impurity in embedding schemes emerging from solid-state physics ), where the coupling with the remaining part of the molecule (the bath) is treated by including only the states, which are strongly entangled with the impurity. The system-bath separation in DMET follows the standard construction recipe of open quantum systems. It is also closely related to the separation of the lattice in each DMRG microiteration step. 

The core of the embedding is a Schmidt decomposition of the total state into many-particle states defined on the system and on the environment
\begin{equation}
 \begin{aligned}
  | \Phi \rangle =& \sum_{ij} C_{ij} | \Phi_i^s \rangle | \Phi_j^e \rangle = \sum_i | C_i \Phi_i^s \rangle \left( \sum_j \frac{C_{ij}}{C_i} | \Phi_j^e \rangle \right) \\
                 =& \sum_i C_i | \Phi_i^s \rangle | \tilde{\Phi}_i^e \rangle \, ,
 \end{aligned}     
 \label{eq:DMET}
\end{equation}
where $| \Phi_i^s \rangle$ and $| \Phi_j^e \rangle$ are many-particle basis states defined on system and environment, respectively. 
In this formal presentation of the decomposition, we hide the environmental degree of freedom $j$ in such a way that every relevant
state $i$ on the system couples to exactly one {\it contracted} basis state on the environment. In other words, the double sum has been replaced
by a single summation, which has the advantage that one requires only one basis state in the environment to couple to each basis state
on the system. However, the contraction over index $j$ highlights that each corresponding state in the environment may be difficult to
construct for the product ansatz to be accurate.

As an embedding approach that naturally follows from open-systems quantum mechanics, an advantage of DMET is that each portion of the molecule can, in principle, be treated at a high-level of theory, and only couplings between different blocks are considered on the mean-field level. A major limitation of DMET is the representation of the bath with a single determinant that is optimized to match the high-level one-particle density matrix (or its diagonal part, as proposed by Scuseria and co-workers\cite{Scuseria2014_DMET-BrokenSymmetry} in the so-called density embedding theory). This could be overcome, in principle, by replacing the HF wave function by an MPS with a low value of $m$. However, this would require the generalization of DMET to post-HF parametrizations of the low-level wave function, which has been proposed recently for some electronic structure methods\cite{VanNeck2017_BlockStates-DMET} but not for the DMRG.

The LAS-SCF method described in Section~\ref{sec:DMRG_Theory} represents a way to embed a CAS-SCF wave function in a CAS-SCF environment. This is realized by expressing the wave function as in Eq.~(\ref{eq:LAS-SCF}), \textit{i.e.} as a direct product of CAS-SCF wave functions localized on different portions of the orbital space. However, the parametrization of Eq.~(\ref{eq:LAS-SCF}) neglects the entanglement between different orbital groups and therefore, as we discussed above, its accuracy will probably strongly dependent on the partition of the orbitals. This is especially true for cases in which this partition is not trivially determined by the molecular topology, as in dimers or in molecular aggregates.

A limitation of DMET is that the partitioning of the orbitals in fragments introduces an unbalanced description of orbitals, the one being in the middle of a fragment being described more accurately than the ones lying on the boundary between two fragments. The bootstrap embedding theory introduced by Van Voorhis and co-workers\cite{VanVoorhis2016_BootstrapEmbedding,VanVoorhis2017_Bootstrap2D} aims at solve this problem by applying DMET to multiple partitions of the orbitals and to constrain the one-particle and on-top density matrices to be the one obtained with a partition, where the orbital is in the middle. Bootstrapping embedding has been first introduced for monodimensional spin chains, for which it is trivial to identify the orbitals that are close to the boundary or in the middle of the fragment. The algorithm has been recently extended to molecular systems\cite{VanVoorhis2019_BootstrapMolecules} for which such identification is not trivial.

\subsection{DMRG for molecular spectroscopy}

Another field of applications of DMRG not yet studied thoroughly enough concerns static and dynamical properties and spectroscopy. For comparatively low energies, molecular properties are obtained as derivatives of the energy with respect to a given perturbation.\cite{Helgaker2012_ReviewProperties} The calculation of first-order properties, such as the electric dipole moment, can be simplified with the Hellmann-Feynmann theorem but the calculation of higher-order properties is less trivial and requires, for single-reference methods, the solution of the so-called coupled-perturbed Hartree-Fock equations. The generalization of linear response theory to DMRG has been proposed under the name of linear-response DMRG (LR-DMRG) and applied to the calculation of both static\cite{Dorando2009_AnalyticalResponseFunction} and dynamical\cite{Nakatani2014_LinearResponseDMRG} optical properties of polyenes.\cite{Dorando2009_AnalyticalResponseFunction} These pilot studies show the way toward further improvements. First, they rely on first-generation formulations of DMRG, but an extension to the MPS/MPO formulation would be possible within the framework introduced recently to solve the time-dependent Schr\"{o}dinger equation with DMRG.\cite{Lubich2014_TimeIntegrationTT} LR-DMRG could also be, in principle, generalized to higher-order properties following, for example, the theory already available for CAS-SCF wave functions.\cite{Olsen1985_MCSCF-LinearResponse} The LR-DMRG theory can also benefit from generalizations proposed for other electronic structure methods including, for example, the damped response formalism,\cite{Kristensen2009_DampedLR-QuasiEnergy,Coriani2012_DampedResponse-CC} which avoids instabilities in the definition of the response function under resonance conditions. Transition properties between electronic states of different multiplicity can be calculated either within a fully-relativistic formulation of DMRG variationally or by optimizing the two target electronic states with non-relativistic DMRG and subsequently calculating transition properties by applying perturbation theory. Within the latter scheme, the two electronic wave functions (i) may be represented within the same set of molecular orbtial set\cite{Chan2016_StateInteraction-ZeroField,Chan2018_gTensors}, but (ii) are usually optimized independently, and therefore, are expressed in terms of different sets of molecular orbitals. In the latter case, the calculation of transition properties between two states then requires a state-interaction algorithm formulated in a biorthonormal basis,\cite{Malmqvist1989_CASSI} originally developed for traditional CAS-SCF and RAS-SCF state interaction (SI). We have extended this scheme toward MPS representations of the two states and introduced MPS state interaction (MPS-SI) in a biorthonormal basis and applied it to the calculation of spin-orbit couplings, $g$ tensors, and zero-field splittings.\cite{Knecht2016_StateInteraction}

We conclude by mentioning a promising application of DMRG to computational spectroscopy, \textit{i.e.} the calculation of X-ray absorption spectra. CAS-SCF and its restricted extension, RAS-SCF, have been applied for the calculation of core excitation energies of transition metal complexes. The necessity of restricting the excitations through RAS arises from the need of directly targeting core excited states, without optimizing all the lower lying ones.\cite{Pinjari2014_RASSCF-LEdge,Ludberg2016_IronXRay} In addition to the usual problem of selecting the CAS, in RAS-SCF the orbitals must be divided into different groups, which may affect the accuracy. DMRG could bypass these problems in two respects. First of all, the energy-specific formulations of DMRG\cite{Dorando2007_TargetingExcitedStates,Yu2017_ShiftAndInvertMPS,Devakul2017,Baiardi2019_HighEnergy-vDMRG} would allow directly targeting excited states, without the need of imposing any restriction on the excitation degree. Furthermore, the \texttt{AutoCAS} algorithm described above could automatize the selection of the active orbitals to be included in the CAS, in this way bypassing the limits of RAS-SCF. The application of DMRG to core excitation energies would also test the reliability of the MPS parametrization to describe highly excited states. As already highlighted above, the efficiency of DMRG should be maximal for many-body localized excited states. This condition should be met for core excitations governed by only a small number of orbitals.

\section{DMRG and selected CI: a possible match?}
\label{sec:CriticalComparison}

In Section~\ref{sec:intro}, we mentioned that DMRG has become, together with selected CI approaches, a state-of-the-art method for large multireference problems. The size of the largest active space targeted by selected CI are (118,32) for the iterative CI scheme by Zimmerman\cite{Ziemmermann2017_iCI} and (76,28) for the heath-bath CI of Sharma.\cite{Sharma2018_Fast-HBCI} For DMRG, the largest calculations reported up to now target active spaces with size (120,77) and (118,55).\cite{Chan2018_Nitrogenase,Chan2019_FeMoCo-DMRG} We will now discuss the factors that impede applications of these methods for larger systems and discuss possible improvements to push them beyond these limits.

All different flavors of selected CI approaches\cite{Malrieu1983_ImprovedCIPSI,Alavi2009_FCIQMCOriginal,HeadGordon2016_SelectedCI,Schriber2016_AdaptiveCI,Eriksen2017_VirtualOrbitalSelectedCI,Ziemmermann2017_iCI} rely on the following full CI expansion

\begin{equation}
  | \Phi \rangle = \sum_{\sigma_1} \ldots \sum_{\sigma_L} C_{\sigma_1,\ldots,\sigma_L} | \sigma_1 \cdots \sigma_L \rangle \, .
  \label{eq:CIExpansion}
\end{equation}

The CI tensor $\bm{C} = \{ C_{\sigma_1,\ldots,\sigma_L} \}$ is in most cases too large to be optimized with standard algorithms and is assumed \textit{a priori} to be sparse. Based on this assumption, different selected CI algorithm differ in the strategy for identifying efficiently the non-zero elements of $\bm{C}$. As we already highlighted, DMRG does not attempt to exploit the sparsity of the CI tensor, but parametrizes it as a TT. The resulting wave function, the MPS, encodes efficiently strong entanglement effects, and this includes both sparse and non-sparse CI expansions.

The linear relation between the occupation number vector basis $| \sigma_1 \cdots \sigma_L \rangle$ and the wave function is the core advantage of selected CI schemes. The representation of the non-relativistic electronic Hamiltonian in this basis is sparse and non-zero matrix elements are easily obtained by applying the well-known Slater-Condon rules. Based on this sparsity assumption and exploiting the fact that the Hamiltonian couples determinants that differ by at most two excitations, the most relevant contributions to Eq.~(\ref{eq:CIExpansion}) can be identified, for example, by exploiting energy estimates obtained from perturbation theory\cite{Schriber2016_AdaptiveCI} or directly from the size of the matrix elements.\cite{Holmes2016_HBCI,Sharma2017_HBCI} The screening can also be performed via a stochastic exploration of the configurational space, a route that is followed in FCIQMC.\cite{Alavi2009_FCIQMCOriginal,Alavi2011_Initiator-FCIQMC,Alavi2012_FCIQMC-Solids} A slightly different scheme is incremental CI\cite{Ziemmermann2017_iCI,Eriksen2017_VirtualOrbitalSelectedCI,Gauss2018_MBE-FCI-WeaklyCorrelated} which approximates the full CI energy with a many-body expansion and, therefore, avoids the construction of the wave function as in Eq.~(\ref{eq:CIExpansion}). A major advantage of selected CI approaches is that the sampling of the CI space is, in most cases, trivially parallelizable. Moreover, second-order perturbative corrections have a rather straightforward expression and can be evaluated either with deterministic\cite{Holmes2016_HBCI} or stochastic algorithms,\cite{Sharma2017_HBCI,Sharma2018_Fast-HBCI} the latter option being more appealing as it can be trivially parallelized.

These advantages are, however, counterbalanced by several limitations connected to the implicit assumption that the tensor $\bm{C}$ of Eq.~(\ref{eq:CIExpansion}) is sparse. Based on these assumptions, any selected CI scheme constructs iteratively the CI expansion with incremental algorithms. However, the degree of sparsity of the CI expansion, and hence the efficiency, decreases for strongly-correlated systems. This phenomenon has been observed for incremental CI,\cite{Gauss2018_MBE-FCI-WeaklyCorrelated} where an incremental expansion is explicitly constructed, but is expected to have a strong impact on other selected CI schemes as well. Even if we assume that the fraction of non-null elements in $\bm{C}$ is a constant independent of system size, the number of non-null elements of $\bm{C}$ will show the same scaling. Most selected CI schemes require to store these elements and, therefore, the computational cost is expected to grow quickly for large systems due to huge memory requirements.

Unlike selected CI, DMRG optimizes a non-linear parametrization of the wave function in terms of the CI basis $| \sigma_1 \cdots \sigma_L \rangle$. The simple Slater-Condon rules do not apply anymore and the calculation of matrix elements of the Hamiltonian involves complex contractions of MPSs with its MPO representation. An efficient calculation of these matrix elements relies on the possibility of storing intermediate contractions between MPSs and MPOs that can be reused within the sweep-based optimization. The memory needed to store these contractions, which is the bottleneck of DMRG in most cases, depends on the length of the DMRG lattice and on the size of MPSs and MPOs. Increasing the size of a CAS clearly affects the first parameter, but has an indirect effect on the other two quantities. For large DMRG lattices long-range Coulomb interactions are represented with a large MPO, whose ground state is encoded, in turn, by a less compact MPS. In practice, owing to all these factors, the memory requirement becomes prohibitive for active spaces with more than 100 orbitals. Despite this memory bottleneck, the scaling of DMRG is formally polynomial in the system size for Hamiltonians that follow the area law. We highlighted in Section~\ref{sec:DMRG_Theory} that this is not the case for quantum-chemical Hamiltonians, neither for the electronic nor for the vibrational ones, for which the bond dimension $m$ may scale exponentially with the system size. Nevertheless, quantum-chemical applications of DMRG showed that the bond dimension depends only weakly on the overall dimension. This suggests that the exponential scaling of $m$ with system size is hardly ever encountered in practice, and that tensor networks provide a significantly more compact \textit{ansatz} than a standard full CI approach. This makes tensor-network approaches a more reliable starting point for the design of new multi-configurational approaches targeting more than, say, 100 orbitals. We highlight that the bond dimension $m$ depends indirectly on the system size, especially in the presence of long-range interactions. For this reason, the MPS parametrization might become less and less convenient when targeting very large systems and more complex parametrizations might become more appealing.

The design of efficient multi-configurational schemes could exploit the possibility of combining the advantages of selected CI and tensor-network approaches. The CI parametrization of Eq.~(\ref{eq:CIExpansion}) that simplifies the calculation of the representation of the Hamiltonian, is intrinsically different compared to an MPS. However, recent studies providing a thorough characterization of the MPS space with concepts taken from differential geometry\cite{Haegeman2011_TDDMRG-MPSMPO,Holtz2012_ManifoldTT,Verstraete2013_GeometryMPS} proved that the set composed of all MPSs can be approximated as a linear subspace of the full configurational space in the vicinity of a reference MPS. Therefore, selected CI calculations may be performed in this linear space. This idea has been already exploited in the context of stochastic perturbation theory\cite{Guo2018_StochasticPT-DMRG} and for the calculation of excitation energies\cite{Wouters2013_ThoulessTheorem-DMRG} with wave functions encoded as MPSs. In addition, this linearized approximation of the MPS space can be sampled with stochastic algorithms with, for example, FCIQMC or heat-bath CI. The combination of DMRG with stochastic methods would pave the route towards a massive parallelization of DMRG that is non-trivial with standard formulations. Similar ideas have been explored only in Ref.~\citenum{Chan2014_AFQMC-MatrixProductStates}, where DMRG is coupled with auxiliary-field quantum Monte Carlo. In this respect, several studies characterizing various tensor factorizations from a mathematical perspective, based on differential geometry concepts, appeared in the literature in the last years.\cite{Verstraete2013_GeometryMPS,Verstraete2016_Gradient-PEPS,Kressner2016_Preconditioner-Tensor,Rakhuba2018_LowRank,Verstraete2019_Tangent-uMPS} These studies could drive the design of new, more efficient tensor networks and of algorithms alternative to DMRG.

We already highlighted that a major advantage of stochastic methods, including FCIQMC\cite{Alavi2009_FCIQMCOriginal} and semistochastic HBCI,\cite{Sharma2017_HBCI-Semistochastic,Sharma2018_Fast-HBCI} is the possibility of a massive parallelization of the critical steps of the algorithm, i.e. the time-evolution of the walkers for FCIQMC and the calculation of the perturbative correction for HBCI. DMRG cannot be parallelized as trivially because the sweep-based optimization is intrinsically sequential. As discussed by Sabzevari and Sharma,\cite{Sharma2018_Fast-VMC} any non-linear wave function parametrization can be optimized stochastically, provided that the overlap of the wave function with a given Slater determinant $| \bm{\sigma} \rangle$ can be calculated efficiently. This is the case of matrix product states, for which, however, ALS is still more efficient than other optimization schemes. For other cases, such as for multi-reference CI\cite{Sharma2019_Stochastic-MRCI} or for symmetry-projected Jastrow mean-field wave functions,\cite{Sharma2019_Projected-VMC} the stochastic optimization can be more efficient, as well as easier to implement, than the deterministic one. The combination of a compact wave function parametrization and a massively parallelizable optimization algorithm could drive the design of new tensor network states that encode efficiently dynamical correlation effects.

\section{Conclusions}
\label{sec:conclusions}

The density matrix renormalization group algorithm is currently one of the reference methods for the calculation of full-CI energies in a space of up to about 100 spatial orbitals. Originally applied for diagonalizing spin Hamiltonians of interest in solid-state physics, we discussed its extension to quantum chemical \textit{ab initio} Hamiltonians. DMRG possesses most of the desirable properties of a reliable electronic structure theory. It is size consistent and corresponds to a well-defined wave function parametrization, the matrix product state. The accuracy and computational cost of DMRG can be controlled by the size of the matrix product state, which is governed by a single parameter, the bond dimension $m$. In the limit of an infinitely large value for the bond dimension, the full CI result is recovered. Converged energies are, however, obtained with compact matrix product states with low values of the bond dimension. Through the fixation of $m$ based on the spectrum of the reduced density matrix, the DMRG wave function becomes self-adaptive to the quantum many-particle structure under consideration.

DMRG belongs to a set of new methods that have emerged in the last years to perform large-scale CI calculations, such as full-CI quantum Monte Carlo,\cite{Alavi2009_FCIQMCOriginal,Alavi2011_Initiator-FCIQMC} or selected CI approaches.\cite{Malrieu1983_ImprovedCIPSI,Schriber2016_AdaptiveCI,Sharma2017_HBCI,Gauss2018_MBE-FCI-WeaklyCorrelated,Ziemmermann2017_iCI} These other approaches are, however, based on a standard full CI wave function and limit the computational cost of the optimization by avoiding the full-dimensional matrix diagonalization. This is a major difference compared to DMRG, which is based on a parametrization radically different from the full CI one and which replaces the diagonalization with an iterative approximation of the full CI wave function.

Already in its original formulations, the good convergence of DMRG makes it considerably more efficient than the majority of other CAS-based approaches. A further increase of efficiency has been achieved by applying strategies borrowed from standard CAS calculations. DMRG calculations are most efficiently performed on orbitals exhibiting strong static correlation. The remaining dynamical correlation can then be included from perturbation theory. The combination of DMRG with perturbation theory has made it one of the \quotes{gold standard} reference methods for multi-configurational molecules. The current main limitation of most DMRG perturbation theory approaches is that the evaluation of sum-over-states expressions does not exploit the compact structure of a matrix product state and requires the calculation of high-order density matrices. For this reason, the evaluation of the perturbative correction is in most cases the main bottleneck of the overall calculation. Expressing the second-order perturbation to the energy as a variational problem\cite{Sinanoglu1961_Hylleraas} enables one to express the first-order correction to a wave function as an MPS and to calculate it with a sweep optimization.\cite{Sharma2014_Hylleraas-DMRG,Sharma2016_QuasiDegenerate-PT,Sharma2017_MRPT-DMRG} However, the resulting MPS has usually a large bond dimension and work is currently in progress to obtain a compact representation for it.\cite{Guo2018_DMRG-Hylleraas,Guo2018_StochasticPT-DMRG} In recent years, cost-effective alternative to standard perturbative approaches have been introduced, such as the driven similarity renormalization group approach\cite{Evangelista2014_DSRG,Evangelista2018_ACI+DSRG} or the generalized RPA scheme,\cite{Pernal2018_AC-GroundExcited,Pernal2018_ExtendedRPACorrelation} and their integration within the DMRG framework would enable one to target even larger active spaces.

The matrix-product-state parametrization has been designed to target Hamiltonians without any long-range interaction. Generalizations of this parametrization, broadly known as tensor network states, have been studied to provide compact representation of wave functions for more general Hamiltonians, comprising both short- and long-ranged interactions. The applications of tensor network states to quantum chemistry has been, however, rather limited due to the lack of efficient general optimization methods.

Although DMRG is a general algorithm that can be applied to the optimization of the ground state of any Hamiltonian it has been mostly applied to electronic structure problems in quantum chemistry. Recent generalizations of DMRG include the extension to vibrational, rotational, and vibronic Hamiltonians with remarkable speed-ups compared to state-of-the-art variational approaches. These results suggest that DMRG is general and robust enough to be successfully applied to other types of Hamiltonians of interest in quantum chemistry. The MPS/MPO-based formalism is the natural framework to be applied to such a variety of systems, since most algorithms developed to construct the MPO representation of an operator require as unique input their second-quantized form\cite{Frowis2010_MPOGeneric,Keller2015_MPSMPODMRG} and, once the MPO is built, the optimization algorithm is the same independent of the Hamiltonian.

The MPS parametrization has been studied in numerical analysis under the name of \quotes{tensor train} factorization and has been applied to a wide range of problems beyond the solution of eigenvalue equations. These recent developments have paved the way for the application of DMRG to the time-dependent Schr\"{o}diger equation. The success of its application to the solution of the time-independent Schr\"{o}dinger equation suggests that DMRG will also become one of the reference methods for large-scale quantum dynamics simulations.

\section*{Acknowledgements}
This work was supported by ETH Zurich (ETH Fellowship No. FEL-49 18-1).

%

\end{document}